%% file: R_LTi_condition_v12.tex
\documentclass[reprint, superscriptaddress, amsmath, amssymb, aps, prl]{revtex4-1}

\include{my_macro}

\usepackage{graphicx}
\usepackage{dcolumn}
\usepackage{bm}
\usepackage{color}
\usepackage{multirow}

\newcommand{\vti}{\ensuremath{v_{{\rm th}i}}}

\newcommand{\Upar}{U_\parallel}
\newcommand{\Uper}{U_\perp}
\newcommand{\Utor}{U_\phi}
\newcommand{\Btor}{B_\phi}
\newcommand{\Bpol}{B_p}
\newcommand{\rhoi}{\rho_i}

\newcommand{\ly}{\ell_y}

\newcommand{\tst}{\tau_{\rm st}}

\newcommand{\gEbar}{\bar\gamma_E}
\newcommand{\Qturb}{\bar Q_{\rm turb}}

\newcommand{\dn}{\delta n}

\newcommand{\LTi}{L_{T_i}}
\newcommand{\RLTi}{R/L_{T_i}}
\newcommand{\RLTic}{R/L_{T_i,c}}
\newcommand{\LTe}{L_{T_e}}

\newcommand{\Ln}{L_n}
\newcommand{\RLTe}{R/L_{T_e}}
\newcommand{\RLn}{R/\Ln}

\newcommand{\nust}{\nu_{*i}}

\renewcommand{\eqref}[1]{Eq.~(\ref{#1})}
\newcommand{\myfig}[3][3.3in]{
\begin{figure}[tbp]
\includegraphics[width=#1]{#2}%
\caption{#3\label{fig:#2}}%
\end{figure}
}
\newcommand{\figref}[1]{Fig.~\ref{fig:#1}}
\newcommand{\tableref}[1]{Table~\ref{#1}}

\begin{document}

\preprint{APS/123-QED}

\title{Local dependence of ion temperature gradient on magnetic configuration, rotational shear and turbulent heat flux in MAST}
\ycghim
\arfield
\aaschekochihin
\eghighcock
\cmichael
\mastteam

\date{\today}

\begin{abstract}
Experimental data from the Mega Amp Spherical Tokamak (MAST) is used to show that the inverse gradient scale length of the ion temperature $\RLTi$ (normalized to the major radius $R$) has its strongest {\em local} correlation with the rotational shear and the pitch angle of the magnetic field (or, equivalently, an inverse correlation with $q/\varepsilon$, the safety factor/the inverse aspect ratio). Furthermore, $\RLTi$ is found to be {\em inversely} correlated with the gyro-Bohm-normalized local turbulent heat flux estimated from the density fluctuation level measured using a 2D Beam Emission Spectroscopy (BES) diagnostic. These results can be explained in terms of the conjecture that the turbulent system adjusts to keep $\RLTi$ close to a certain critical value (marginal for the excitation of turbulence) determined by local equilibrium parameters (although not necessarily by linear stability).
\end{abstract}

\maketitle

\paragraph{Introduction.}
A key physics challenge posed by magnetically confined plasmas in fusion devices is how their internal energy can be kept from being transported too fast from the core to the periphery. The problem is primarily one of turbulent transport, the temperature gradient between the edge and the core of a toroidal plasma supplying the free energy for the turbulent fluctuations that then enhance the effective thermal diffusivity and relax the gradient. It is the {\em ion} temperature gradient (ITG) that is expected to cause the most virulent instabilities (on ion Larmor scales) \cite{rudakov_doklady_1961,coppi_pof_1967,cowley_pfb_1991} and then to be self-consistently limited by the resulting turbulence \cite{horton_rmp_1999}. If we view the edge ion temperature as fixed by the physics and engineering aspects of the tokamak design that will not concern us here \cite{snyder_nf_2011}, the key question is how to maximize the ion temperature gradient. We therefore wish to inquire, experimentally, what this gradient depends on and how. Motivated by the fact (or the conjecture) that the ion-scale turbulence is largely determined by the {\em local} (to a given flux surface) equilibrium conditions \cite{candy_pop_2004,candy_pop_2009,barnes_pop_2010,abel_rpp_2012,ghim_prl_2012} and in turn acts back to adjust them {\em locally}, we ask what local parameters are most strongly correlated with the corresponding value of $\RLTi$, the inverse radial gradient scale length of the ion temperature ($\LTi^{-1}=\labs\partial\ln T_i/\partial r\rabs$) normalized to the major radius $R$ of the torus.
 \newline\indent
How universal are any such measured dependences likely to be for situations with different global conditions, e.g., different neutral-beam-injection (NBI) heating powers? It has been recognized for some time that the turbulent heat flux tends to increase very strongly (much faster than linearly) with $\RLTi$, a phenomenon known as ``stiff'' transport \cite{kotschenreuther_pop_1995,barnes_prl_2011_107,mantica_prl_2009,mantica_prl_2011,mantica_ppcf_2011}. If (or when) the transport is stiff, any experimentally measured relationship between $\RLTi$ and other equilibrium parameters should be quite close to some critical manifold in the parameter space separating dominant turbulent transport from a non-turbulent or weakly turbulent state (the ``zero-turbulence manifold'' \cite{highcock_prl_2012}). This critical manifold would be independent of the power input and can be represented as a local parameter dependence of the critical temperature gradient $\RLTic = f(q,\varepsilon,\hat s,\Utor',\RLn,\RLTe,\nu_{ii},T_i/T_e,\beta_i,\dots)$, where $q$ is the safety factor (number of toroidal revolutions per one poloidal revolution of the magnetic field around the torus on a given flux surface), $\varepsilon=r/R$ the inverse aspect ratio ($r$ is the minor radius of the flux surface), $\hat s=\partial\ln q/\partial\ln r$ the magnetic shear, $\Utor'=\partial\Utor/\partial r$ the radial shear of the mean toroidal rotation velocity $\Utor$, $\Ln$ and $\LTe$ the gradient scale lengths of the plasma density and electron temperature, $\nu_{ii}$ the ion collision rate, $T_i/T_e$ the ion-to-electron temperature ratio, $\beta_i=8\pi nT_i/B^2$ the ion-to-magnetic pressure ratio and ``$\dots$'' stand for everything else (e.g., the many parameters required to fully describe the magnetic configuration) \footnote{We do not suggest {\em causality} between all these parameters and $\RLTi$ --- all local equilibrium characteristics, including $\RLTi$, jointly adjust to form the critical manifold.}.  We stress that $\RLTic$ need not be the same as the threshold for the existence of linearly unstable eigenmodes. Two known examples when it is not are the ``Dimits upshift'' of $\RLTi$ above the linear stability threshold \cite{dimits_pop_2000,rogers_prl_2000,mikkelsen_prl_2008} and the case of sufficiently large $\Utor'$ when the system is linearly stable but strong transient excitations \cite{newton_ppcf_2010,schekochihin_ppcf_2012} lead to sustained subcritical turbulence \cite{barnes_prl_2011,highcock_prl_2010,highcock_pop_2011}. Thus, in general, there is a nonlinear threshold with some definite dependence on local equilibrium parameters. 
\newline\indent
Recent theoretical \cite{newton_ppcf_2010,schekochihin_ppcf_2012} and numerical \cite{highcock_prl_2012} investigations suggest that $q/\varepsilon$ and $\Utor'$ may be the most important such parameters, at least at low $\hat s$. Let us explain why this is. It is well known, both from experimental measurements \cite{ghim_ppcf_2012,field_ppcf_2009} and theory \cite{hinton_pf_1985,cowley_clr_1986,abel_rpp_2012}, that strong (finite-Mach) flows in a tokamak are predominantly toroidal (certainly when plasma is heated by tangential neutral beams, which produce a toroidal torque) \footnote{Any mean poloidal flow exceeding the diamagnetic velocity would be  damped by collisions \cite{connor_ppcf_1987,catto_pf_1987}.}. Therefore, any radial shear in the toroidal flow results in sheared flow in both the perpendicular ($\Uper' = \lp\Bpol/B\rp \Utor'$, $\Bpol$ is the poloidal field) and parallel ($\Upar' = \lp\Btor/B\rp \Utor'$, $\Btor$ is the toroidal field) directions. While perpendicular flow shear is known (theoretically \cite{dorland_ppcnfr_1994,waltz_pop_1994,dimits_nf_2001,kinsey_pop_2005,camenen_pop_2009,roach_ppcf_2009,casson_pop_2009,barnes_prl_2011,highcock_prl_2010} and experimentally \cite{burrell_pop_1997,burrell_pop_1999,mantica_prl_2009,mantica_prl_2011,mantica_ppcf_2011,schaffner_prl_2012}) to suppress turbulence and the associated transport, parallel flow shear can drive turbulence via the ``parallel-velocity-gradient'' (PVG) instability \cite{catto_pf_1973,newton_ppcf_2010,schekochihin_ppcf_2012}. The average ratio of these two shearing rates on a flux surface, $\Upar'/\Uper' = \Btor/\Bpol$, can be approximated by $q/\varepsilon$ and so the degree to which sheared equilibrium flow suppresses or drives turbulence is expected to depend on this parameter. Indeed, numerical studies of ITG- and PVG-driven turbulence have shown that the critical $\RLTic$ at any given $\Utor'$ increases with decreasing $q/\varepsilon$ (at least for low $\hat s$ \cite{highcock_prl_2012}); while at any given $q/\varepsilon$, $\RLTic$ increases with increasing $\Utor'$ provided the latter is not too large \cite{barnes_prl_2011,highcock_prl_2010,highcock_pop_2011,highcock_prl_2012} (as in most real tokamaks). 
\newline\indent
A comprehensive numerical parameter scan of the dependence of $\RLTic$ on all other potentially important local quantities ($\hat s$, $\RLn$, $\RLTe$, $\nu_{ii}$, $T_i/T_e$, $\beta_i$, etc.) is probably unaffordable in the near future. Faster progress can be made experimentally. In this Letter, our first goal is to establish, based on a relatively sizable dataset for MAST, what the most important parameters for the critical manifold are: we will show that, indeed, the local value of $\RLTi$ is most strongly correlated inversely with the local $q/\varepsilon$ and positively with the local rotational shear --- consistently with the result obtained in \cite{highcock_prl_2012}.
\newline\indent
Our second goal is to obtain an experimental signature that the measured $\RLTi$ is determined by --- or, more precisely, correlated with --- the local characteristics of the ion-scale turbulence, directly measured by the 2D beam emission spectroscopy (BES) diagnostic \cite{field_rsi_2012}. We will show that not only does a strong correlation between $\RLTi$ and an estimated turbulent heat flux level exist but its (at the first glance, counterintuitive) inverse nature is consistent with $\RLTi$ staying close to the critical threshold $\RLTic$ and hence with stiff transport. 

\paragraph{Equilibrium parameters.} 
A database was compiled of equilibrium quantities (and turbulence characteristics; see below) from 39 neutral-beam-heated discharges from the 2011 MAST experimental campaign. These discharges had a double-null diverted (DND) magnetic configuration, no pellet injection and no applied resonant magnetic perturbations. Mean electron density $n_e$ and temperature $T_e$ were measured with the Thomson scattering system \cite{scannell_rsi_2010}, mean impurity ion (C$^{6+}$) temperature $T_i$ and the toroidal flow velocity $\Utor$ with the Charge eXchange Recombination Spectroscopy (CXRS) system \cite{conway_rsi_2006} (we assumed that in these discharges the impurity and bulk ions have negligible differences in their temperature and flow velocities \cite{kim_pfb_1990}).  The local magnetic pitch angle ($\Bpol/\Btor$) was measured with the Motional Stark Effect (MSE) diagnostic \cite{debock_rsi_2008}; pressure- and MSE-constrained \texttt{EFIT} equilibria \cite{lao_nf_1985} were used to obtain the field strength $B$. All parameters were determined over $5\:ms$ intervals either by averaging if the diagnostic's temporal resolution was smaller or by interpolation if it was larger than $5\:ms$. Only data points from a limited range of minor radii $0.6<r/a<0.7$ ($r=a$ is the edge of the plasma) were used, in order to minimize any correlations between various quantities due to their profile dependence alone (thus, we did not attempt to {\em prove} locality here; see, however, \cite{ghim_prl_2012}). In total, 988 data points were available (out of which 109 points are from H-mode discharges).
\newline\indent
From this information, we constructed 7 local dimensionless parameters, which, motivated by theoretical models and common sense, we deemed {\em a priori} the most important ones (we also give the range of variation of each parameter): $\RLTi \in [0.08,20.3]$, $q/\varepsilon\in[4.0,16.3]$, $\hat s \in [1.2,6.0]$, $\gEbar \equiv \Uper'\tst=\pi r\Utor'/\vti \in [0.005,2.5]$, $\RLn \in [0.04,13.8]$, $\RLTe \in [1.43,22.7]$, $\nust\equiv \nu_{ii}\tst\in[0.003,0.12]$, $T_i/T_e \in [0.5,1.7]$. The ion collision rate $\nu_{ii}$ and the perpendicular velocity shear $\Uper'$ (which is used instead of $\Utor'$) were normalized, as in \cite{ghim_prl_2012}, to the ion parallel streaming time $\tst=\Lambda/\vti$, where $\vti=\sqrt{2T_i/m_i}$ and $\Lambda=\pi r B/\Bpol$ is the connection length (the approximate distance along the field line from the outboard to the inboard side of the torus, expected to determine the parallel correlation scale of the turbulence \cite{barnes_prl_2011_107}; if the flux surfaces had been circular, $\Lambda\approx \pi qR$). The local magnetic configuration is represented by $q/\varepsilon$ and $\hat s$. The choice of $q/\varepsilon$ was motivated by the physical considerations outlined in the Introduction; since $\varepsilon$ varied little in our database, we cannot distinguish any individual correlations of $\RLTi$ with $q$ and $\varepsilon$. It is left for further study whether other properties of the flux surfaces matter (e.g., Shafranov shift, triangularity, elongation, etc.; some of these may, in fact, affect the stiff-transport threshold \cite{beer_pop_1997,mikkelsen_prl_2008}). We have not included $n$, $T_i$, $T_e$, which are not normalizable by any natural local quantities; note that $\RLTi$ usually has a large but trivial correlation with $T_i$: larger temperature gradients lead to larger temperatures in the core. We also have excluded $\beta_i=8\pi nT_i/B^2$ because, in the absence of large variation of $B$ in our dataset, $\beta_i$ is simply the normalized ion pressure and, similarly to $T_i$, has a large positive correlation with $\RLTi$ (it remains to be investigated whether larger magnetic fluctuations at larger $\beta_i$ are large enough to have a nontrivial effect on turbulent transport \cite{rechester_prl_1978,pueschel_pop_2010,nevins_prl_2011,hatch_prl_2012,guttenfelder_pop_2012,doerk_pop_2012,abel_njp_2012}). 

\paragraph{Correlation analysis.} We perform a Canonical Correlation Analysis (CCA) \cite{hotelling_bio_1936} with $\RLTi$ treated as the dependent variable and the other 7 local parameters itemized above as independent ones. This amounts to finding the maximum correlations between  $\ln(\RLTi)$ and linear combinations of logarithms of 1, 2, 3, \dots, or 7 
other parameters, leading to an effective statistical dependence 
\begin{equation}
\label{eq:cca_fit}
\frac{R}{\LTi} = \lp\frac{q}{\varepsilon}\rp^{\alpha_1}\gEbar^{\alpha_2}\nust^{\alpha_3} 
 \lp\frac{R}{\LTe}\rp^{\alpha_4}\hat s^{\alpha_5}\lp\frac{R}{\Ln}\rp^{\alpha_6}\lp\frac{T_i}{T_e}\rp^{\alpha_7}.
\end{equation} 
This is of course not valid if the dependence of $\RLTi$ on any of the parameters is non-monotonic. A non-monotonic dependence on $\gEbar$ is, in fact, expected, with $\RLTi$ first increasing, then decreasing at larger values of $\gEbar$ due to increased transport from the PVG-driven turbulence \cite{barnes_prl_2011,highcock_prl_2010,highcock_pop_2011,highcock_prl_2012}. However, the range of values of $\gEbar$ in our database do not extend to sufficiently high values for such a dependence to be observed (see \figref{q_eps_tau_st_tau_sh_R_LTi}). 
\newline\indent
The results are shown in \tableref{table:cca_coeff}, where the values of the canonical correlation (i.e., the correlation coefficient between the logarithms of $\RLTi$ and the right-hand of \eqref{eq:cca_fit}) are given together with the corresponding exponents $\alpha_1,\dots,\alpha_7$. We start by calculating the individual correlations of $\RLTi$ with each of the 7 parameters and then include pairs, triplets, etc., only if the correlation improves. We see that the strongest individual correlation of $\RLTi$ are with $q/\varepsilon$ (61\%) and $\gEbar$ (46\%). The overall fit is measurably improved (66\%) if both are included. Including further parameters does not make a significant difference; the third strongest (although not very strong) dependence is on $\nust$. 

\begin{table}[t]\caption{Results of CCA performed assuming \eqref{eq:cca_fit}. 
Wherever $0$ appears, that means that the CCA was performed without including the corresponding parameter.} 
\label{table:cca_coeff}
\begin{tabular}{c | c c c c c c c } 
Canonical  & $q/\varepsilon$ & $\gEbar$ & $\nust$ & $\RLTe$ & $\hat s$ & $\RLn$ & $T_i/T_e$   \\
correlation & $\alpha_1$ & $\alpha_2 $ & $\alpha_3 $ & $\alpha_4 $ & $\alpha_5 $ & $\alpha_6$ & $\alpha_7$ \\
\hline
2.7\% & 0 & 0 & 0 & 0 & 0 & 0 &$-3.1$ \\
13.5\% & 0 & 0 & 0 & 0 & 0 & $-0.83$ & 0 \\
15.4\% & 0 & 0 & 0 & 0 & $-3.41$ & 0 & 0 \\
16.8\% & 0 & 0 & 0 & $-2.3$ & 0 & 0 & 0 \\
36\% & 0 & 0 & $-0.93$ & 0 & 0 & 0 & 0 \\
46\% & 0 & $0.94$ & 0 & 0 & 0 & 0 & 0 \\
61\% & $-1.69$ & 0 & 0 & 0 & 0 & 0 & 0 \\
\hline
62\% & $-1.50$ & 0 & $-0.21$ & 0 & 0 & 0 & 0 \\
66\% & $-1.30$ & $0.40$ & 0 & 0 & 0 & 0 & 0 \\
\hline
67\% & $-1.19$ & $0.38$ & $-0.15$ & 0 & 0 & 0 & 0 \\
\hline
69\% & $-1.12$ & $0.37$ & $-0.19$ & $-0.27$ & $0.21$ & $-0.09$ & $-0.51$ \\
\hline
\end{tabular}
\end{table}

\myfig[3.25in]{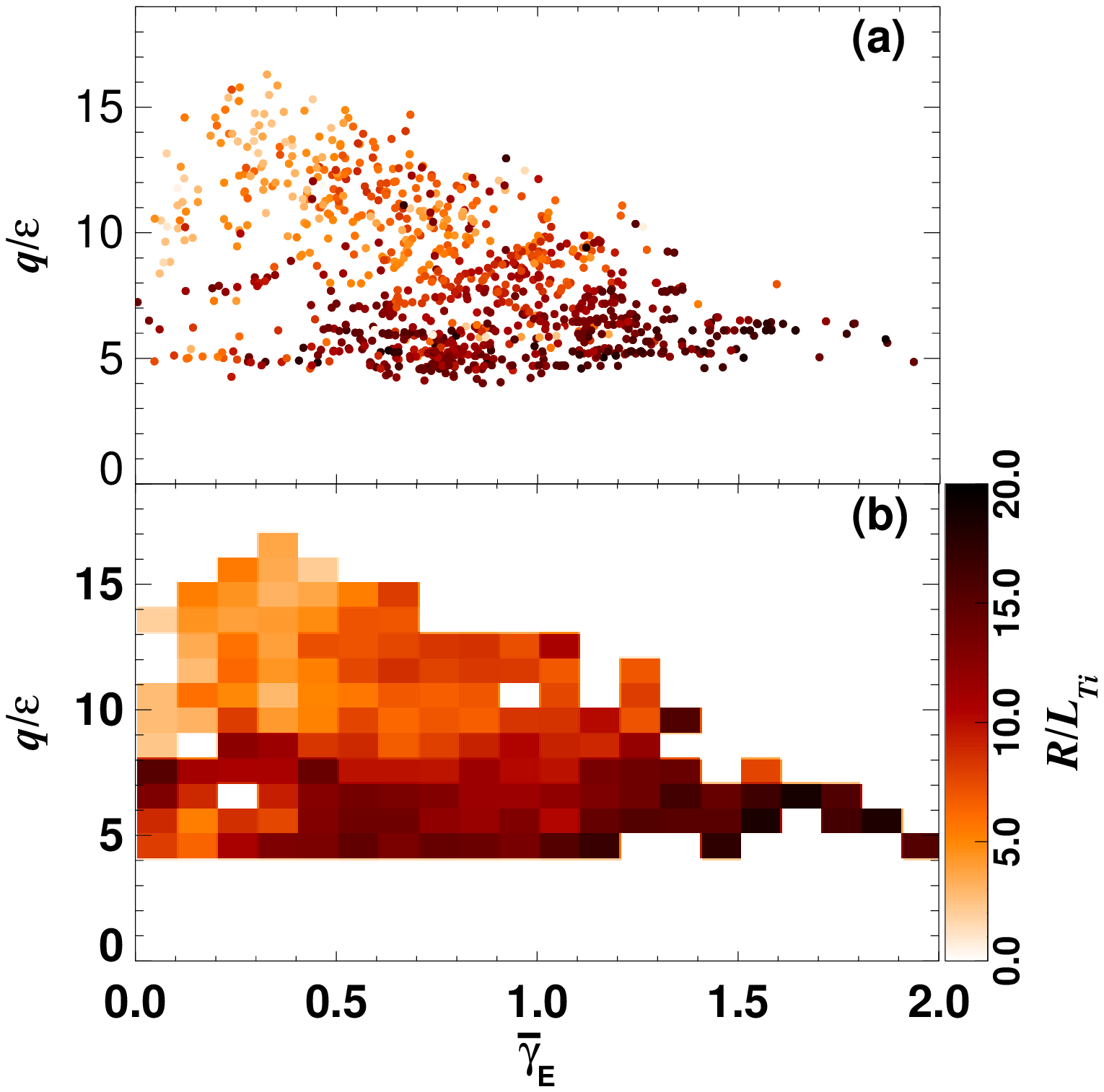}{The dependence of $\RLTi$ (color) on $q/\varepsilon$ and $\gEbar = \pi r\Utor'/\vti$, showing (a) the individual data points and (b) mean values of $\RLTi$ within rectangular bins.}
The dependence of $\RLTi$ on\ $q/\varepsilon$ and $\gEbar$ is shown in \figref{q_eps_tau_st_tau_sh_R_LTi}. $\RLTi$ generally increases with decreasing $q/\varepsilon$ and increasing $\gEbar$ \footnote{The broad scatter of data points in \figref{q_eps_tau_st_tau_sh_R_LTi}(a) suggests that the correlation between $q/\varepsilon$ and $\gEbar$ is weak; the lack of higher values of $\gEbar$ at large $q/\varepsilon$ is due to the fact that the flow shear is weak at earlier times in the discharges, when the central value of $q$ is high.}, broadly consistent with the expectations based on intuitive physical reasoning (explained in the Introduction) and on the numerical study of \cite{highcock_prl_2012}. 
\newline\indent
The conclusion is that, at least on a very rough qualitative level, it is sensible to consider $\RLTi$ to be a function primarily of $q/\varepsilon$ and $\gEbar$. Since the $q$ profile tends to change more slowly in tokamaks than other equilibrium profiles \footnote{It can be proven that the functional dependence $q(\psi)$, where $\psi$ is the flux-surface label, only changes on the resistive timescale of the mean magnetic field \cite{abel_njp_2012}.}, it may be useful to think of a critical curve $\RLTic(\gEbar)$ \cite{parra_prl_2011} parametrized by $q/\varepsilon$ \cite{highcock_prl_2012}, the latter quantity containing the essential information about the nature of the magnetic cage confining the plasma. 
 
\paragraph{Collisionality dependence.}
Even though the $\nust$ dependence of $\RLTi$ is not as strong as $q/\varepsilon$ and $\gEbar$, some discernible inverse correlation between $\RLTi$ and $\nust$ might be expected because zonal flows, believed to suppress turbulence \cite{rogers_prl_2000,diamond_ppcfreview_2005}, should be more strongly damped at higher ion collisionality \cite{hinton_ppcf_1999,lin_prl_1999,ricci_prl_2006,xiao_pop_2007}. To isolate this dependence, we selected data points for approximately fixed $\gEbar\in[0.7,0.8]$ and $q/\varepsilon\in[5,6]$  (the largest number of data points could be found within these narrow ranges and no measurable correlation between $\RLTi$ and $q/\varepsilon$ or $\gEbar$ was present). The resulting \figref{R_LTi_others} confirms a degree of inverse correlation between $\RLTi$ and $\nust$. 

\paragraph{Turbulent heat flux.}
The turbulent ion heat flux through a given flux surface is (very approximately!) $Q_i\sim n T_i\chi_i/\LTi$, where the effective turbulent diffusivity $\chi_i\sim \delta u^2\tau_c$, $\delta u\sim (c/B)\varphi/\ly$ is the (radial) fluctuating $\mathbf{E}\times\mathbf{B}$ velocity, $\tau_c$ its correlation time, $\ly$ its poloidal correlation scale and $\varphi$ the fluctuating electrostatic potential. The latter can be estimated from density fluctuations using the approximation of Boltzmann electrons: $e\varphi/T_e\approx\dn/n$ ($e$ is the proton charge, $n$ and $\dn$ the mean and fluctuating density, respectively). Both theory of ITG turbulence \cite{barnes_prl_2011_107} and the BES measurements in MAST \cite{ghim_prl_2012} suggest that $\tau_c\sim\tau_{*i}=\ly\LTi/\vti\rhoi$ (the drift time; $\rhoi$ is the ion Larmor radius). Collecting all this together, we estimate the gyro-Bohm-normalized turbulent ion heat flux:
\begin{equation}
\frac{Q_{i,\rm turb}}{Q_{\rm gB}}\sim 
\frac{\rhoi}{\ly}\lp\frac{R}{\rhoi}\rp^2\lp\frac{T_e}{T_i}\frac{\dn}{n}\rp^2\equiv\Qturb,
\label{eq:Q}
\end{equation}
where $Q_{\rm gB} = n T_i \vti \rhoi^2 / R^2$.
\newline\indent
Since ion-scale density fluctuations in MAST can be measured directly by the BES system, $\Qturb$ can be obtained independently of any transport reconstruction models such as TRANSP \cite{hawryluk_TRANSP_1980}. The method of determining $\dn/n$ and $\ly$ using the BES system on MAST (8 radial $\times$ 4 vertical channels with spatial resolution of $\approx 2$~cm \cite{field_rsi_2012}) is explained in detail in \cite{ghim_prl_2012}. This is done from the covariance and correlation functions of the photon intensity fluctuations, averaged over the same $5\:ms$ intervals for the same 39 discharges as the equilibrium quantities studied above, although not in all intervals there was good BES data. Restricted to the radial range $0.6<r/a<0.7$, the number of available data points for $\dn/n$ and $\ly$ was 102. 

\myfig[3.25in]{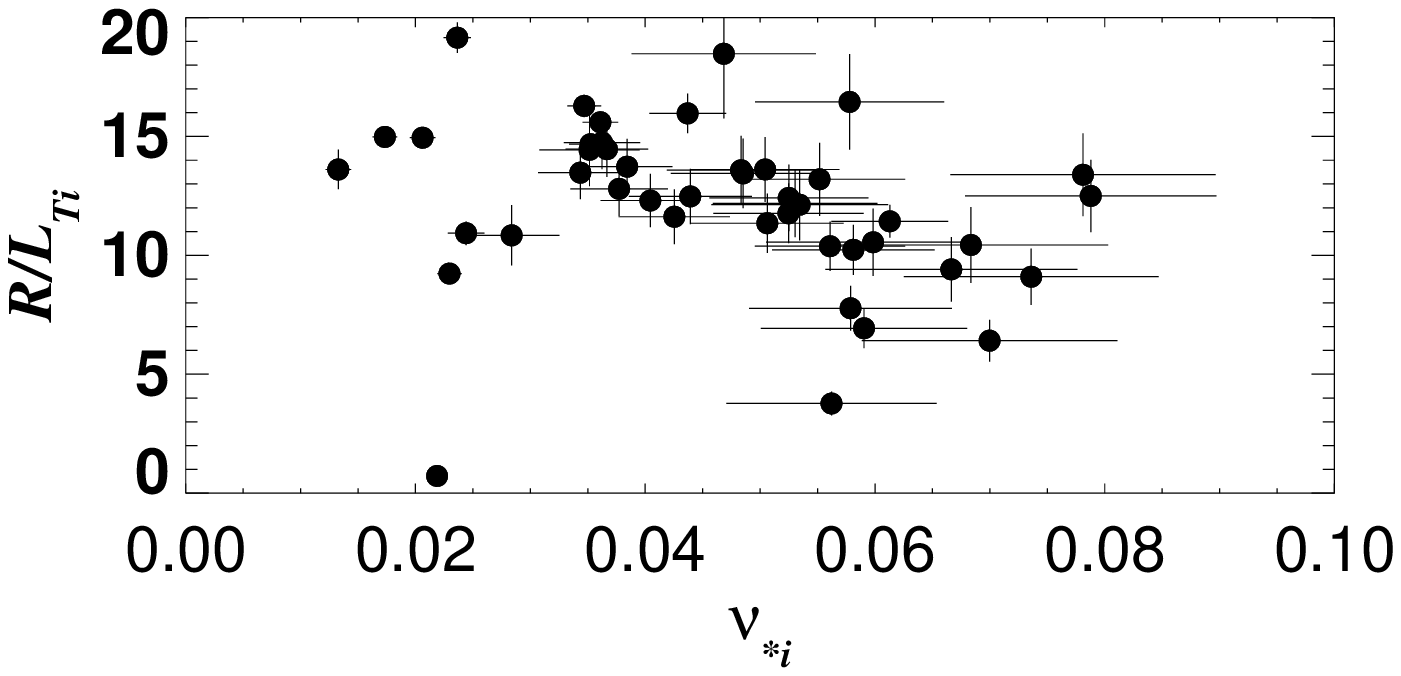}{$\RLTi$ vs.~$\nust$ for fixed $\gEbar\in[0.7,0.8]$ and $q/\varepsilon\in[5,6]$.}

\paragraph{Inverse correlation between $\RLTi$ and $\Qturb$.} It is shown in \figref{qi_plots}(a) ($\Qturb$ vs.\ $q/\varepsilon$ and $\gEbar$; cf.\ \figref{q_eps_tau_st_tau_sh_R_LTi}) and \figref{qi_plots}(b) ($\Qturb$ vs.\ $\RLTi$) that smaller $\Qturb$ is observed where $\RLTi$ is large and vice versa. This is perhaps counterintuitive as one might expect that the larger $\RLTi$, the more turbulent the plasma and so the larger the turbulent heat flux. This would indeed have been the case had $\RLTi$ been externally fixed, as in local flux-tube simulations, where $\Qturb$ does increase with $\RLTi$ \cite{dimits_pop_2000,barnes_prl_2011_107}. In contrast, in the real plasma, a certain amount of power flows through a flux surface and, if transport is stiff, the temperature gradient (along with other equilibrium quantities) adjusts to stay close to the critical gradient defined by the manifold $\RLTic(\gEbar,q/\varepsilon)$. Indeed, in plasmas with both power and momentum injection, there is a regime with $\RLTi$ close to $\RLTic$ where the turbulent and the neoclassical (collisional) transport are comparable and in which a larger heat flux results in lower $\RLTi$. For a detailed explanation, we refer the reader to \cite{parra_prl_2011} (see also Fig.~4(b) of \cite{highcock_prl_2010}, which should be compared to our \figref{qi_plots}(b)). In brief, in the neoclassical regime, the momentum transport is much less efficient than the heat transport, while in the turbulent regime, they are comparable (the turbulent Prandtl number is order unity \cite{casson_pop_2009,barnes_prl_2011,highcock_pop_2011, meyer_nf_2009} while the neoclassical one is small \cite{hinton_pf_1985}), so, as a larger heat flux takes the system (slightly) farther from the marginal state, this leads to much more efficient momentum transport, hence smaller velocity shear $\gEbar$, hence a regime with less suppression of turbulence and smaller $\RLTic$ (see \figref{qi_plots}(c), where the correspondence between larger $\Qturb$, lower $\RLTi$ and lower $\gEbar$ is shown at approximately fixed $q/\varepsilon\in [9,11]$; cf.\ Fig.~3(b) of \cite{parra_prl_2011}). We stress that all of this happens quite close to marginality and so our experimental observation of an inverse correlation between $\RLTi$ and $\Qturb$ provides circumstantial evidence that $\RLTi$ in MAST is indeed close to its critical value \footnote{This situation is distinct from the experiments on transport stiffness on JET \cite{mantica_prl_2009,mantica_prl_2011,mantica_ppcf_2011} in that they provided vigorous extra heating power using localized ion-cyclotron-resonance heating (ICRH) to depart far from marginality.}. 

\myfig[3in]{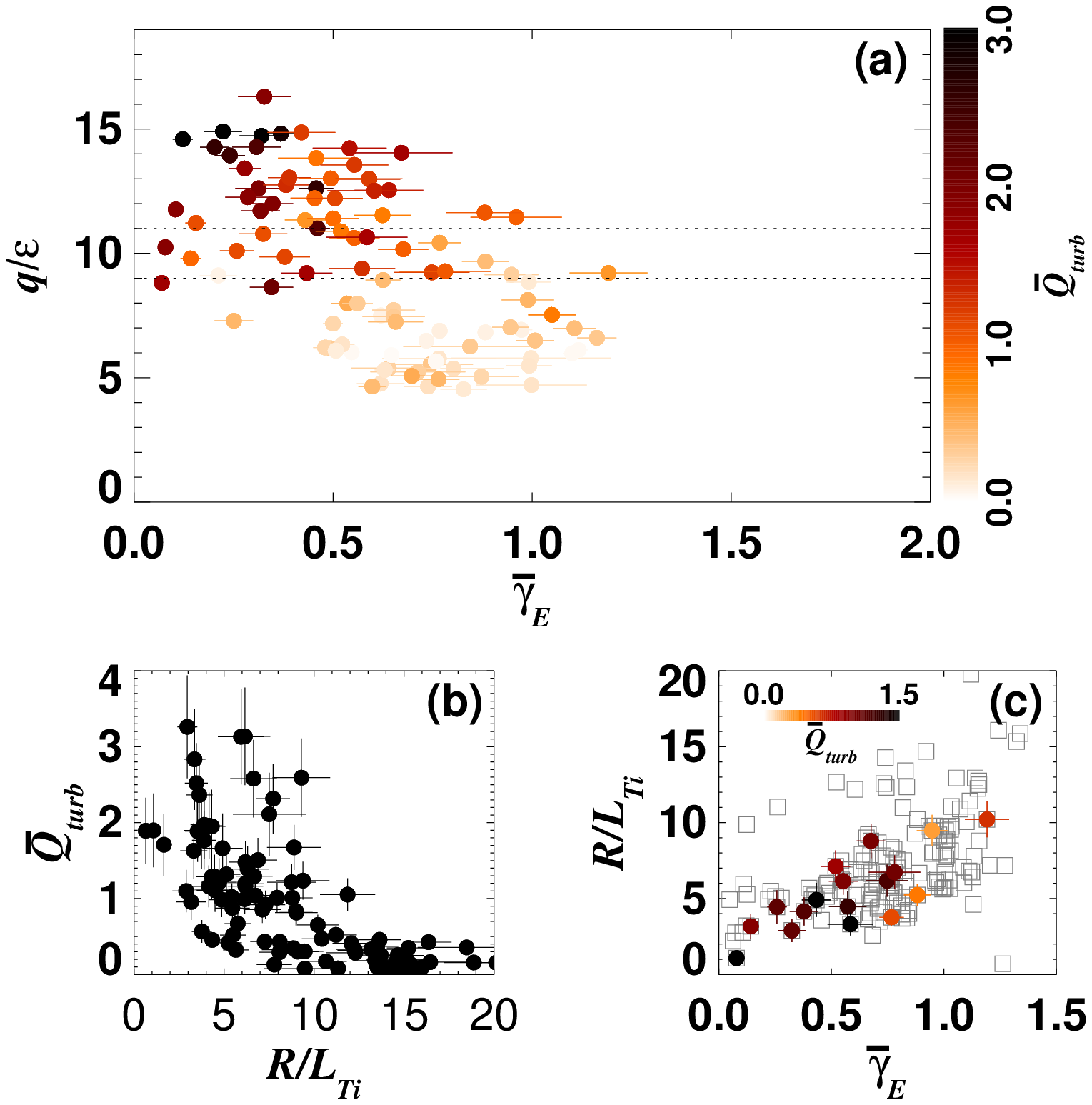}{(a) $\Qturb$ (color) calculated from BES data according to \eqref{eq:Q} vs.\ $q/\varepsilon$ and $\gEbar$ (cf.\ \figref{q_eps_tau_st_tau_sh_R_LTi}); (b) $\Qturb$ vs.\ $\RLTi$(cf.\ Fig.~4(b) of \cite{highcock_prl_2010}); (c) $\Qturb$ (color) vs.\ $\RLTi$ and $\gEbar$ for a fixed range of $q/\varepsilon\in[9,11]$, indicated by the dotted horizontal lines in (a); open squares are data from \figref{q_eps_tau_st_tau_sh_R_LTi}(a) for which BES measurements were not available (cf.\ Fig.~3(b) of \cite{parra_prl_2011}).} 

\paragraph{Conclusion.}
We have found that the normalized inverse ion-temperature-gradient scale length $\RLTi$ has its strongest local correlation with $q/\varepsilon$ and the shear in the equilibrium toroidal flow: $\RLTi$ increases with increasing shear, which is a well known effect, and with decreasing $q/\varepsilon$, which corresponds to an increasing ratio of the perpendicular to the parallel shearing rates. We note that a similar dependence of $\RLTi \lp q/\varepsilon, \gEbar\rp$ is also observed in JET \cite{fern_unpub_2012}, suggesting that the inverse correlation between $\RLTi$ and $q/\varepsilon$ is perhaps ubiquitous, as would be the case if $\RLTi$ were generally fixed at some locally determined critical value \cite{highcock_prl_2012}. Furthermore, we have found an {\em inverse} correlation between $\RLTi$ and the gyro-Bohm-normalized turbulent heat flux (estimated via direct measurements of density fluctuations) and argued that this is consistent with $\RLTi$ always remaining close to a critical manifold $\RLTic\lp q/\varepsilon, \gEbar\rp$ separating the turbulent and non-turbulent regimes \cite{highcock_prl_2010,parra_prl_2011,highcock_pop_2011} (stiff transport). It is thus plausible that we have essentially produced this critical manifold for the MAST discharges we investigated. Practically, our results suggest that $\RLTi$ can be increased by lowering the $q/\varepsilon$, which is relatively easier and less expensive than increasing the shearing rate in tokamak operations \footnote{With tangential NBI heating, it is difficult to increase the toroidal Mach number --- and hence the equilibrium flow shear --- because of the fixed ratio of injected torque to power at a fixed injection energy.}. 

\begin{acknowledgments}
We thank I.~Abel, M.~Barnes, J.~Connor, S.~Cowley, W.~Dorland, D.~Dunai, R.~Fern, G.~Hammett, T.~Horbury, F.~I.~Parra, J.~B.~Taylor, M.~Valovi\v{c} and S.~Zoletnik for valuable discussions. This work was supported by the RCUK Energy Programme under grant EP/I501045, the Kwanjeong Educational Foundation (Y-cG), the European Communities under the contract of Association between EURATOM and CCFE (Y-cG, ARF) and by the Leverhulme Trust International Network for Magnetised Plasma Turbulence. The views and opinions expressed herein do not necessarily reflect those of the European Commission.
\end{acknowledgments}


\bibliography{reference}

\end{document}

%% file: my_macro.tex
\newcommand{\lp}{\left(}
\newcommand{\rp}{\right)}

\newcommand{\labs}{\left|}
\newcommand{\rabs}{\right|}

\newcommand{\rudolphpeierls}{Rudolf Peierls Centre for Theoretical Physics, University of Oxford, Oxford, OX1 3NP, UK}

\newcommand{\culham}{EURATOM/CCFE Fusion Association, Culham Science Centre, Abingdon, OX14 3DB, UK}

\newcommand{\mertoncollege}{Merton College, Oxford, OX1 4JD, UK}
\newcommand{\magdalen}{Magdalen College, Oxford, OX1 4AU, UK}

\newcommand{\anu}{Research School of Physical Sciences and Engineering, Australian National University, Canberra, ACT 0200, Australia}

\newcommand{\ycghim}{\author{Y.-c.\ Ghim}\email{y.kim1@physics.ox.ac.uk}\affiliation{\rudolphpeierls}\affiliation{\culham}}
\newcommand{\aaschekochihin}{\author{A.\ A.\ Schekochihin}\affiliation{\rudolphpeierls}\affiliation{\mertoncollege}}
\newcommand{\arfield}{\author{A.\ R.\ Field}\affiliation{\culham}}
\newcommand{\cmichael}{\author{C.\ Michael}\affiliation{\anu}\affiliation{\culham}}

\newcommand{\mastteam}{\author{the\ MAST\ Team}\affiliation{\culham}}
\newcommand{\eghighcock}{\author{E. G.\ Highcock}\affiliation{\magdalen}\affiliation{\rudolphpeierls}}

%% file: R_LTi_condition_v12.bbl
\begin{thebibliography}{71}%
\makeatletter
\providecommand \@ifxundefined [1]{%
 \@ifx{#1\undefined}
}%
\providecommand \@ifnum [1]{%
 \ifnum #1\expandafter \@firstoftwo
 \else \expandafter \@secondoftwo
 \fi
}%
\providecommand \@ifx [1]{%
 \ifx #1\expandafter \@firstoftwo
 \else \expandafter \@secondoftwo
 \fi
}%
\providecommand \natexlab [1]{#1}%
\providecommand \enquote  [1]{``#1''}%
\providecommand \bibnamefont  [1]{#1}%
\providecommand \bibfnamefont [1]{#1}%
\providecommand \citenamefont [1]{#1}%
\providecommand \href@noop [0]{\@secondoftwo}%
\providecommand \href [0]{\begingroup \@sanitize@url \@href}%
\providecommand \@href[1]{\@@startlink{#1}\@@href}%
\providecommand \@@href[1]{\endgroup#1\@@endlink}%
\providecommand \@sanitize@url [0]{\catcode `\\12\catcode `\$12\catcode
  `\&12\catcode `\#12\catcode `\^12\catcode `\_12\catcode `\%12\relax}%
\providecommand \@@startlink[1]{}%
\providecommand \@@endlink[0]{}%
\providecommand \url  [0]{\begingroup\@sanitize@url \@url }%
\providecommand \@url [1]{\endgroup\@href {#1}{\urlprefix }}%
\providecommand \urlprefix  [0]{URL }%
\providecommand \Eprint [0]{\href }%
\providecommand \doibase [0]{http://dx.doi.org/}%
\providecommand \selectlanguage [0]{\@gobble}%
\providecommand \bibinfo  [0]{\@secondoftwo}%
\providecommand \bibfield  [0]{\@secondoftwo}%
\providecommand \translation [1]{[#1]}%
\providecommand \BibitemOpen [0]{}%
\providecommand \bibitemStop [0]{}%
\providecommand \bibitemNoStop [0]{.\EOS\space}%
\providecommand \EOS [0]{\spacefactor3000\relax}%
\providecommand \BibitemShut  [1]{\csname bibitem#1\endcsname}%
\let\auto@bib@innerbib\@empty
\bibitem [{\citenamefont {Rudakov}\ and\ \citenamefont
  {Sagdeev}(1961)}]{rudakov_doklady_1961}%
  \BibitemOpen
  \bibfield  {author} {\bibinfo {author} {\bibfnamefont {L.~I.}\ \bibnamefont
  {Rudakov}}\ and\ \bibinfo {author} {\bibfnamefont {R.~Z.}\ \bibnamefont
  {Sagdeev}},\ }\href@noop {} {\bibfield  {journal} {\bibinfo  {journal} {Dokl.
  Akad. Nauk SSSR}\ }\textbf {\bibinfo {volume} {138}},\ \bibinfo {pages} {581}
  (\bibinfo {year} {1961})}\BibitemShut {NoStop}%
\bibitem [{\citenamefont {{Coppi}}\ \emph {et~al.}(1967)\citenamefont
  {{Coppi}}, \citenamefont {{Rosenbluth}},\ and\ \citenamefont
  {{Sagdeev}}}]{coppi_pof_1967}%
  \BibitemOpen
  \bibfield  {author} {\bibinfo {author} {\bibfnamefont {B.}~\bibnamefont
  {{Coppi}}}, \bibinfo {author} {\bibfnamefont {M.~N.}\ \bibnamefont
  {{Rosenbluth}}}, \ and\ \bibinfo {author} {\bibfnamefont {R.~Z.}\
  \bibnamefont {{Sagdeev}}},\ }\href {\doibase 10.1063/1.1762151} {\bibfield
  {journal} {\bibinfo  {journal} {Phys. Fluids}\ }\textbf {\bibinfo {volume}
  {10}},\ \bibinfo {pages} {582} (\bibinfo {year} {1967})}\BibitemShut
  {NoStop}%
\bibitem [{\citenamefont {{Cowley}}\ \emph {et~al.}(1991)\citenamefont
  {{Cowley}}, \citenamefont {{Kulsrud}},\ and\ \citenamefont
  {{Sudan}}}]{cowley_pfb_1991}%
  \BibitemOpen
  \bibfield  {author} {\bibinfo {author} {\bibfnamefont {S.~C.}\ \bibnamefont
  {{Cowley}}}, \bibinfo {author} {\bibfnamefont {R.~M.}\ \bibnamefont
  {{Kulsrud}}}, \ and\ \bibinfo {author} {\bibfnamefont {R.}~\bibnamefont
  {{Sudan}}},\ }\href {\doibase 10.1063/1.859913} {\bibfield  {journal}
  {\bibinfo  {journal} {Phys. Fluids B}\ }\textbf {\bibinfo {volume} {3}},\
  \bibinfo {pages} {2767} (\bibinfo {year} {1991})}\BibitemShut {NoStop}%
\bibitem [{\citenamefont {{Horton}}(1999)}]{horton_rmp_1999}%
  \BibitemOpen
  \bibfield  {author} {\bibinfo {author} {\bibfnamefont {W.}~\bibnamefont
  {{Horton}}},\ }\href {\doibase 10.1103/RevModPhys.71.735} {\bibfield
  {journal} {\bibinfo  {journal} {Rev. Mod. Phys.}\ }\textbf {\bibinfo {volume}
  {71}},\ \bibinfo {pages} {735} (\bibinfo {year} {1999})}\BibitemShut
  {NoStop}%
\bibitem [{\citenamefont {{Snyder}}\ \emph {et~al.}(2011)\citenamefont
  {{Snyder}}, \citenamefont {{Groebner}}, \citenamefont {{Hughes}},
  \citenamefont {{Osborne}}, \citenamefont {{Beurskens}}, \citenamefont
  {{Leonard}}, \citenamefont {{Wilson}},\ and\ \citenamefont
  {{Xu}}}]{snyder_nf_2011}%
  \BibitemOpen
  \bibfield  {author} {\bibinfo {author} {\bibfnamefont {P.~B.}\ \bibnamefont
  {{Snyder}}}, \bibinfo {author} {\bibfnamefont {R.~J.}\ \bibnamefont
  {{Groebner}}}, \bibinfo {author} {\bibfnamefont {J.~W.}\ \bibnamefont
  {{Hughes}}}, \bibinfo {author} {\bibfnamefont {T.~H.}\ \bibnamefont
  {{Osborne}}}, \bibinfo {author} {\bibfnamefont {M.}~\bibnamefont
  {{Beurskens}}}, \bibinfo {author} {\bibfnamefont {A.~W.}\ \bibnamefont
  {{Leonard}}}, \bibinfo {author} {\bibfnamefont {H.~R.}\ \bibnamefont
  {{Wilson}}}, \ and\ \bibinfo {author} {\bibfnamefont {X.~Q.}\ \bibnamefont
  {{Xu}}},\ }\href {\doibase 10.1088/0029-5515/51/10/103016} {\bibfield
  {journal} {\bibinfo  {journal} {Nucl. Fusion}\ }\textbf {\bibinfo {volume}
  {51}},\ \bibinfo {pages} {103016} (\bibinfo {year} {2011})}\BibitemShut
  {NoStop}%
\bibitem [{\citenamefont {{Candy}}\ \emph {et~al.}(2004)\citenamefont
  {{Candy}}, \citenamefont {{Waltz}},\ and\ \citenamefont
  {{Dorland}}}]{candy_pop_2004}%
  \BibitemOpen
  \bibfield  {author} {\bibinfo {author} {\bibfnamefont {J.}~\bibnamefont
  {{Candy}}}, \bibinfo {author} {\bibfnamefont {R.~E.}\ \bibnamefont
  {{Waltz}}}, \ and\ \bibinfo {author} {\bibfnamefont {W.}~\bibnamefont
  {{Dorland}}},\ }\href {\doibase 10.1063/1.1695358} {\bibfield  {journal}
  {\bibinfo  {journal} {Phys. Plasmas}\ }\textbf {\bibinfo {volume} {11}},\
  \bibinfo {pages} {L25} (\bibinfo {year} {2004})}\BibitemShut {NoStop}%
\bibitem [{\citenamefont {{Candy}}\ \emph {et~al.}(2009)\citenamefont
  {{Candy}}, \citenamefont {{Holland}}, \citenamefont {{Waltz}}, \citenamefont
  {{Fahey}},\ and\ \citenamefont {{Belli}}}]{candy_pop_2009}%
  \BibitemOpen
  \bibfield  {author} {\bibinfo {author} {\bibfnamefont {J.}~\bibnamefont
  {{Candy}}}, \bibinfo {author} {\bibfnamefont {C.}~\bibnamefont {{Holland}}},
  \bibinfo {author} {\bibfnamefont {R.~E.}\ \bibnamefont {{Waltz}}}, \bibinfo
  {author} {\bibfnamefont {M.~R.}\ \bibnamefont {{Fahey}}}, \ and\ \bibinfo
  {author} {\bibfnamefont {E.}~\bibnamefont {{Belli}}},\ }\href {\doibase
  10.1063/1.3167820} {\bibfield  {journal} {\bibinfo  {journal} {Phys.
  Plasmas}\ }\textbf {\bibinfo {volume} {16}},\ \bibinfo {pages} {060704}
  (\bibinfo {year} {2009})}\BibitemShut {NoStop}%
\bibitem [{\citenamefont {{Barnes}}\ \emph {et~al.}(2010)\citenamefont
  {{Barnes}}, \citenamefont {{Abel}}, \citenamefont {{Dorland}}, \citenamefont
  {{G{\"o}rler}}, \citenamefont {{Hammett}},\ and\ \citenamefont
  {{Jenko}}}]{barnes_pop_2010}%
  \BibitemOpen
  \bibfield  {author} {\bibinfo {author} {\bibfnamefont {M.}~\bibnamefont
  {{Barnes}}}, \bibinfo {author} {\bibfnamefont {I.~G.}\ \bibnamefont
  {{Abel}}}, \bibinfo {author} {\bibfnamefont {W.}~\bibnamefont {{Dorland}}},
  \bibinfo {author} {\bibfnamefont {T.}~\bibnamefont {{G{\"o}rler}}}, \bibinfo
  {author} {\bibfnamefont {G.~W.}\ \bibnamefont {{Hammett}}}, \ and\ \bibinfo
  {author} {\bibfnamefont {F.}~\bibnamefont {{Jenko}}},\ }\href {\doibase
  10.1063/1.3323082} {\bibfield  {journal} {\bibinfo  {journal} {Phys.
  Plasmas}\ }\textbf {\bibinfo {volume} {17}},\ \bibinfo {pages} {056109}
  (\bibinfo {year} {2010})}\BibitemShut {NoStop}%
\bibitem [{\citenamefont {{Abel}}\ \emph {et~al.}(2012)\citenamefont {{Abel}},
  \citenamefont {{Plunk}}, \citenamefont {{Wang}}, \citenamefont {{Barnes}},
  \citenamefont {{Cowley}}, \citenamefont {{Dorland}},\ and\ \citenamefont
  {{Schekochihin}}}]{abel_rpp_2012}%
  \BibitemOpen
  \bibfield  {author} {\bibinfo {author} {\bibfnamefont {I.~G.}\ \bibnamefont
  {{Abel}}}, \bibinfo {author} {\bibfnamefont {G.~G.}\ \bibnamefont {{Plunk}}},
  \bibinfo {author} {\bibfnamefont {E.}~\bibnamefont {{Wang}}}, \bibinfo
  {author} {\bibfnamefont {M.}~\bibnamefont {{Barnes}}}, \bibinfo {author}
  {\bibfnamefont {S.~C.}\ \bibnamefont {{Cowley}}}, \bibinfo {author}
  {\bibfnamefont {W.}~\bibnamefont {{Dorland}}}, \ and\ \bibinfo {author}
  {\bibfnamefont {A.~A.}\ \bibnamefont {{Schekochihin}}},\ }\href@noop {}
  {\bibfield  {journal} {\bibinfo  {journal} {submitted to Rep. Prog. Phys.
  [arXiv:1209.4782]}\ } (\bibinfo {year} {2012})}\BibitemShut {NoStop}%
\bibitem [{\citenamefont {{Ghim}}\ \emph {et~al.}(2012)\citenamefont {{Ghim}},
  \citenamefont {{Schekochihin}}, \citenamefont {{Field}}, \citenamefont
  {{Abel}}, \citenamefont {{Barnes}}, \citenamefont {{Colyer}}, \citenamefont
  {{Cowley}}, \citenamefont {{Parra}}, \citenamefont {{Dunai}}, \citenamefont
  {{Zoletnik}},\ and\ \citenamefont {{the MAST Team}}}]{ghim_prl_2012}%
  \BibitemOpen
  \bibfield  {author} {\bibinfo {author} {\bibfnamefont {Y.-c.}\ \bibnamefont
  {{Ghim}}}, \bibinfo {author} {\bibfnamefont {A.~A.}\ \bibnamefont
  {{Schekochihin}}}, \bibinfo {author} {\bibfnamefont {A.~R.}\ \bibnamefont
  {{Field}}}, \bibinfo {author} {\bibfnamefont {I.~G.}\ \bibnamefont {{Abel}}},
  \bibinfo {author} {\bibfnamefont {M.}~\bibnamefont {{Barnes}}}, \bibinfo
  {author} {\bibfnamefont {G.}~\bibnamefont {{Colyer}}}, \bibinfo {author}
  {\bibfnamefont {S.~C.}\ \bibnamefont {{Cowley}}}, \bibinfo {author}
  {\bibfnamefont {F.~I.}\ \bibnamefont {{Parra}}}, \bibinfo {author}
  {\bibfnamefont {D.}~\bibnamefont {{Dunai}}}, \bibinfo {author} {\bibfnamefont
  {S.}~\bibnamefont {{Zoletnik}}}, \ and\ \bibinfo {author} {\bibnamefont {{the
  MAST Team}}},\ }\href@noop {} {\bibfield  {journal} {\bibinfo  {journal}
  {submitted to Phys. Rev. Lett. [arXiv:1208.5970]}\ } (\bibinfo {year}
  {2012})}\BibitemShut {NoStop}%
\bibitem [{\citenamefont {{Kotschenreuther}}\ \emph {et~al.}(1995)\citenamefont
  {{Kotschenreuther}}, \citenamefont {{Dorland}}, \citenamefont {{Beer}},\ and\
  \citenamefont {{Hammett}}}]{kotschenreuther_pop_1995}%
  \BibitemOpen
  \bibfield  {author} {\bibinfo {author} {\bibfnamefont {M.}~\bibnamefont
  {{Kotschenreuther}}}, \bibinfo {author} {\bibfnamefont {W.}~\bibnamefont
  {{Dorland}}}, \bibinfo {author} {\bibfnamefont {M.~A.}\ \bibnamefont
  {{Beer}}}, \ and\ \bibinfo {author} {\bibfnamefont {G.~W.}\ \bibnamefont
  {{Hammett}}},\ }\href {\doibase 10.1063/1.871261} {\bibfield  {journal}
  {\bibinfo  {journal} {Phys. Plasmas}\ }\textbf {\bibinfo {volume} {2}},\
  \bibinfo {pages} {2381} (\bibinfo {year} {1995})}\BibitemShut {NoStop}%
\bibitem [{\citenamefont {Barnes}\ \emph
  {et~al.}(2011{\natexlab{a}})\citenamefont {Barnes}, \citenamefont {Parra},\
  and\ \citenamefont {Schekochihin}}]{barnes_prl_2011_107}%
  \BibitemOpen
  \bibfield  {author} {\bibinfo {author} {\bibfnamefont {M.}~\bibnamefont
  {Barnes}}, \bibinfo {author} {\bibfnamefont {F.~I.}\ \bibnamefont {Parra}}, \
  and\ \bibinfo {author} {\bibfnamefont {A.~A.}\ \bibnamefont {Schekochihin}},\
  }\href@noop {} {\bibfield  {journal} {\bibinfo  {journal} {Phys. Rev. Lett.}\
  }\textbf {\bibinfo {volume} {107}},\ \bibinfo {pages} {115003} (\bibinfo
  {year} {2011}{\natexlab{a}})}\BibitemShut {NoStop}%
\bibitem [{\citenamefont {{Mantica}}\ \emph {et~al.}(2009)\citenamefont
  {{Mantica}}, \citenamefont {{Strintzi}}, \citenamefont {{Tala}},
  \citenamefont {{Giroud}}, \citenamefont {{Johnson}}, \citenamefont
  {{Leggate}}, \citenamefont {{Lerche}}, \citenamefont {{Loarer}},
  \citenamefont {{Peeters}}, \citenamefont {{Salmi}}, \citenamefont
  {{Sharapov}}, \citenamefont {{van Eester}}, \citenamefont {{de Vries}},
  \citenamefont {{Zabeo}},\ and\ \citenamefont {{Zastrow}}}]{mantica_prl_2009}%
  \BibitemOpen
  \bibfield  {author} {\bibinfo {author} {\bibfnamefont {P.}~\bibnamefont
  {{Mantica}}}, \bibinfo {author} {\bibfnamefont {D.}~\bibnamefont
  {{Strintzi}}}, \bibinfo {author} {\bibfnamefont {T.}~\bibnamefont {{Tala}}},
  \bibinfo {author} {\bibfnamefont {C.}~\bibnamefont {{Giroud}}}, \bibinfo
  {author} {\bibfnamefont {T.}~\bibnamefont {{Johnson}}}, \bibinfo {author}
  {\bibfnamefont {H.}~\bibnamefont {{Leggate}}}, \bibinfo {author}
  {\bibfnamefont {E.}~\bibnamefont {{Lerche}}}, \bibinfo {author}
  {\bibfnamefont {T.}~\bibnamefont {{Loarer}}}, \bibinfo {author}
  {\bibfnamefont {A.~G.}\ \bibnamefont {{Peeters}}}, \bibinfo {author}
  {\bibfnamefont {A.}~\bibnamefont {{Salmi}}}, \bibinfo {author} {\bibfnamefont
  {S.}~\bibnamefont {{Sharapov}}}, \bibinfo {author} {\bibfnamefont
  {D.}~\bibnamefont {{van Eester}}}, \bibinfo {author} {\bibfnamefont {P.~C.}\
  \bibnamefont {{de Vries}}}, \bibinfo {author} {\bibfnamefont
  {L.}~\bibnamefont {{Zabeo}}}, \ and\ \bibinfo {author} {\bibfnamefont
  {K.-D.}\ \bibnamefont {{Zastrow}}},\ }\href {\doibase
  10.1103/PhysRevLett.102.175002} {\bibfield  {journal} {\bibinfo  {journal}
  {Phys. Rev. Lett.}\ }\textbf {\bibinfo {volume} {102}},\ \bibinfo {eid}
  {175002} (\bibinfo {year} {2009})}\BibitemShut {NoStop}%
\bibitem [{\citenamefont {Mantica}\ \emph {et~al.}(2011)\citenamefont
  {Mantica}, \citenamefont {Angioni}, \citenamefont {Challis}, \citenamefont
  {Colyer}, \citenamefont {Frassinetti}, \citenamefont {Hawkes}, \citenamefont
  {Johnson}, \citenamefont {Tsalas}, \citenamefont {deVries}, \citenamefont
  {Weiland}, \citenamefont {Baiocchi}, \citenamefont {Beurskens}, \citenamefont
  {Figueiredo}, \citenamefont {Giroud}, \citenamefont {Hobirk}, \citenamefont
  {Joffrin}, \citenamefont {Lerche}, \citenamefont {Naulin}, \citenamefont
  {Peeters}, \citenamefont {Salmi}, \citenamefont {Sozzi}, \citenamefont
  {Strintzi}, \citenamefont {Staebler}, \citenamefont {Tala}, \citenamefont
  {Van~Eester},\ and\ \citenamefont {Versloot}}]{mantica_prl_2011}%
  \BibitemOpen
  \bibfield  {author} {\bibinfo {author} {\bibfnamefont {P.}~\bibnamefont
  {Mantica}}, \bibinfo {author} {\bibfnamefont {C.}~\bibnamefont {Angioni}},
  \bibinfo {author} {\bibfnamefont {C.}~\bibnamefont {Challis}}, \bibinfo
  {author} {\bibfnamefont {G.}~\bibnamefont {Colyer}}, \bibinfo {author}
  {\bibfnamefont {L.}~\bibnamefont {Frassinetti}}, \bibinfo {author}
  {\bibfnamefont {N.}~\bibnamefont {Hawkes}}, \bibinfo {author} {\bibfnamefont
  {T.}~\bibnamefont {Johnson}}, \bibinfo {author} {\bibfnamefont
  {M.}~\bibnamefont {Tsalas}}, \bibinfo {author} {\bibfnamefont {P.~C.}\
  \bibnamefont {deVries}}, \bibinfo {author} {\bibfnamefont {J.}~\bibnamefont
  {Weiland}}, \bibinfo {author} {\bibfnamefont {B.}~\bibnamefont {Baiocchi}},
  \bibinfo {author} {\bibfnamefont {M.~N.~A.}\ \bibnamefont {Beurskens}},
  \bibinfo {author} {\bibfnamefont {A.~C.~A.}\ \bibnamefont {Figueiredo}},
  \bibinfo {author} {\bibfnamefont {C.}~\bibnamefont {Giroud}}, \bibinfo
  {author} {\bibfnamefont {J.}~\bibnamefont {Hobirk}}, \bibinfo {author}
  {\bibfnamefont {E.}~\bibnamefont {Joffrin}}, \bibinfo {author} {\bibfnamefont
  {E.}~\bibnamefont {Lerche}}, \bibinfo {author} {\bibfnamefont
  {V.}~\bibnamefont {Naulin}}, \bibinfo {author} {\bibfnamefont {A.~G.}\
  \bibnamefont {Peeters}}, \bibinfo {author} {\bibfnamefont {A.}~\bibnamefont
  {Salmi}}, \bibinfo {author} {\bibfnamefont {C.}~\bibnamefont {Sozzi}},
  \bibinfo {author} {\bibfnamefont {D.}~\bibnamefont {Strintzi}}, \bibinfo
  {author} {\bibfnamefont {G.}~\bibnamefont {Staebler}}, \bibinfo {author}
  {\bibfnamefont {T.}~\bibnamefont {Tala}}, \bibinfo {author} {\bibfnamefont
  {D.}~\bibnamefont {Van~Eester}}, \ and\ \bibinfo {author} {\bibfnamefont
  {T.}~\bibnamefont {Versloot}},\ }\href@noop {} {\bibfield  {journal}
  {\bibinfo  {journal} {Phys. Rev. Lett.}\ }\textbf {\bibinfo {volume} {107}},\
  \bibinfo {pages} {135004} (\bibinfo {year} {2011})}\BibitemShut {NoStop}%
\bibitem [{\citenamefont {{Mantica}}\ \emph {et~al.}(2011)\citenamefont
  {{Mantica}}, \citenamefont {{Angioni}}, \citenamefont {{Baiocchi}},
  \citenamefont {{Baruzzo}}, \citenamefont {{Beurskens}}, \citenamefont
  {{Bizarro}}, \citenamefont {{Budny}}, \citenamefont {{Buratti}},
  \citenamefont {{Casati}}, \citenamefont {{Challis}}, \citenamefont
  {{Citrin}}, \citenamefont {{Colyer}}, \citenamefont {{Crisanti}},
  \citenamefont {{Figueiredo}}, \citenamefont {{Frassinetti}}, \citenamefont
  {{Giroud}}, \citenamefont {{Hawkes}}, \citenamefont {{Hobirk}}, \citenamefont
  {{Joffrin}}, \citenamefont {{Johnson}}, \citenamefont {{Lerche}},
  \citenamefont {{Migliano}}, \citenamefont {{Naulin}}, \citenamefont
  {{Peeters}}, \citenamefont {{Rewoldt}}, \citenamefont {{Ryter}},
  \citenamefont {{Salmi}}, \citenamefont {{Sartori}}, \citenamefont {{Sozzi}},
  \citenamefont {{Staebler}}, \citenamefont {{Strintzi}}, \citenamefont
  {{Tala}}, \citenamefont {{Tsalas}}, \citenamefont {{Van Eester}},
  \citenamefont {{Versloot}}, \citenamefont {{deVries}}, \citenamefont
  {{Weiland}},\ and\ \citenamefont {{EFDA Contributors}}}]{mantica_ppcf_2011}%
  \BibitemOpen
  \bibfield  {author} {\bibinfo {author} {\bibfnamefont {P.}~\bibnamefont
  {{Mantica}}}, \bibinfo {author} {\bibfnamefont {C.}~\bibnamefont
  {{Angioni}}}, \bibinfo {author} {\bibfnamefont {B.}~\bibnamefont
  {{Baiocchi}}}, \bibinfo {author} {\bibfnamefont {M.}~\bibnamefont
  {{Baruzzo}}}, \bibinfo {author} {\bibfnamefont {M.~N.~A.}\ \bibnamefont
  {{Beurskens}}}, \bibinfo {author} {\bibfnamefont {J.~P.~S.}\ \bibnamefont
  {{Bizarro}}}, \bibinfo {author} {\bibfnamefont {R.~V.}\ \bibnamefont
  {{Budny}}}, \bibinfo {author} {\bibfnamefont {P.}~\bibnamefont {{Buratti}}},
  \bibinfo {author} {\bibfnamefont {A.}~\bibnamefont {{Casati}}}, \bibinfo
  {author} {\bibfnamefont {C.}~\bibnamefont {{Challis}}}, \bibinfo {author}
  {\bibfnamefont {J.}~\bibnamefont {{Citrin}}}, \bibinfo {author}
  {\bibfnamefont {G.}~\bibnamefont {{Colyer}}}, \bibinfo {author}
  {\bibfnamefont {F.}~\bibnamefont {{Crisanti}}}, \bibinfo {author}
  {\bibfnamefont {A.~C.~A.}\ \bibnamefont {{Figueiredo}}}, \bibinfo {author}
  {\bibfnamefont {L.}~\bibnamefont {{Frassinetti}}}, \bibinfo {author}
  {\bibfnamefont {C.}~\bibnamefont {{Giroud}}}, \bibinfo {author}
  {\bibfnamefont {N.}~\bibnamefont {{Hawkes}}}, \bibinfo {author}
  {\bibfnamefont {J.}~\bibnamefont {{Hobirk}}}, \bibinfo {author}
  {\bibfnamefont {E.}~\bibnamefont {{Joffrin}}}, \bibinfo {author}
  {\bibfnamefont {T.}~\bibnamefont {{Johnson}}}, \bibinfo {author}
  {\bibfnamefont {E.}~\bibnamefont {{Lerche}}}, \bibinfo {author}
  {\bibfnamefont {P.}~\bibnamefont {{Migliano}}}, \bibinfo {author}
  {\bibfnamefont {V.}~\bibnamefont {{Naulin}}}, \bibinfo {author}
  {\bibfnamefont {A.~G.}\ \bibnamefont {{Peeters}}}, \bibinfo {author}
  {\bibfnamefont {G.}~\bibnamefont {{Rewoldt}}}, \bibinfo {author}
  {\bibfnamefont {F.}~\bibnamefont {{Ryter}}}, \bibinfo {author} {\bibfnamefont
  {A.}~\bibnamefont {{Salmi}}}, \bibinfo {author} {\bibfnamefont
  {R.}~\bibnamefont {{Sartori}}}, \bibinfo {author} {\bibfnamefont
  {C.}~\bibnamefont {{Sozzi}}}, \bibinfo {author} {\bibfnamefont
  {G.}~\bibnamefont {{Staebler}}}, \bibinfo {author} {\bibfnamefont
  {D.}~\bibnamefont {{Strintzi}}}, \bibinfo {author} {\bibfnamefont
  {T.}~\bibnamefont {{Tala}}}, \bibinfo {author} {\bibfnamefont
  {M.}~\bibnamefont {{Tsalas}}}, \bibinfo {author} {\bibfnamefont
  {D.}~\bibnamefont {{Van Eester}}}, \bibinfo {author} {\bibfnamefont
  {T.}~\bibnamefont {{Versloot}}}, \bibinfo {author} {\bibfnamefont {P.~C.}\
  \bibnamefont {{deVries}}}, \bibinfo {author} {\bibfnamefont {J.}~\bibnamefont
  {{Weiland}}}, \ and\ \bibinfo {author} {\bibfnamefont {J.}~\bibnamefont
  {{EFDA Contributors}}},\ }\href {\doibase 10.1088/0741-3335/53/12/124033}
  {\bibfield  {journal} {\bibinfo  {journal} {Plasma Phys. Control. Fusion}\
  }\textbf {\bibinfo {volume} {53}},\ \bibinfo {pages} {124033} (\bibinfo
  {year} {2011})}\BibitemShut {NoStop}%
\bibitem [{\citenamefont {{Highcock}}\ \emph {et~al.}(2012)\citenamefont
  {{Highcock}}, \citenamefont {{Schekochihin}}, \citenamefont {{Cowley}},
  \citenamefont {{Barnes}}, \citenamefont {{Parra}}, \citenamefont {{Roach}},\
  and\ \citenamefont {{Dorland}}}]{highcock_prl_2012}%
  \BibitemOpen
  \bibfield  {author} {\bibinfo {author} {\bibfnamefont {E.~G.}\ \bibnamefont
  {{Highcock}}}, \bibinfo {author} {\bibfnamefont {A.~A.}\ \bibnamefont
  {{Schekochihin}}}, \bibinfo {author} {\bibfnamefont {S.~C.}\ \bibnamefont
  {{Cowley}}}, \bibinfo {author} {\bibfnamefont {M.}~\bibnamefont {{Barnes}}},
  \bibinfo {author} {\bibfnamefont {F.~I.}\ \bibnamefont {{Parra}}}, \bibinfo
  {author} {\bibfnamefont {C.~M.}\ \bibnamefont {{Roach}}}, \ and\ \bibinfo
  {author} {\bibfnamefont {W.}~\bibnamefont {{Dorland}}},\ }\href@noop {}
  {\bibfield  {journal} {\bibinfo  {journal} {submitted to Phys. Rev. Lett.
  [arXiv:1203.6455]}\ } (\bibinfo {year} {2012})}\BibitemShut {NoStop}%
\bibitem [{Note1()}]{Note1}%
  \BibitemOpen
  \bibinfo {note} {We do not suggest {\protect \em causality} between all these
  parameters and $R/L_{T_i}$ --- all local equilibrium characteristics,
  including $R/L_{T_i}$, jointly adjust to form the critical
  manifold.}\BibitemShut {Stop}%
\bibitem [{\citenamefont {{Dimits}}\ \emph {et~al.}(2000)\citenamefont
  {{Dimits}}, \citenamefont {{Bateman}}, \citenamefont {{Beer}}, \citenamefont
  {{Cohen}}, \citenamefont {{Dorland}}, \citenamefont {{Hammett}},
  \citenamefont {{Kim}}, \citenamefont {{Kinsey}}, \citenamefont
  {{Kotschenreuther}}, \citenamefont {{Kritz}}, \citenamefont {{Lao}},
  \citenamefont {{Mandrekas}}, \citenamefont {{Nevins}}, \citenamefont
  {{Parker}}, \citenamefont {{Redd}}, \citenamefont {{Shumaker}}, \citenamefont
  {{Sydora}},\ and\ \citenamefont {{Weiland}}}]{dimits_pop_2000}%
  \BibitemOpen
  \bibfield  {author} {\bibinfo {author} {\bibfnamefont {A.~M.}\ \bibnamefont
  {{Dimits}}}, \bibinfo {author} {\bibfnamefont {G.}~\bibnamefont {{Bateman}}},
  \bibinfo {author} {\bibfnamefont {M.~A.}\ \bibnamefont {{Beer}}}, \bibinfo
  {author} {\bibfnamefont {B.~I.}\ \bibnamefont {{Cohen}}}, \bibinfo {author}
  {\bibfnamefont {W.}~\bibnamefont {{Dorland}}}, \bibinfo {author}
  {\bibfnamefont {G.~W.}\ \bibnamefont {{Hammett}}}, \bibinfo {author}
  {\bibfnamefont {C.}~\bibnamefont {{Kim}}}, \bibinfo {author} {\bibfnamefont
  {J.~E.}\ \bibnamefont {{Kinsey}}}, \bibinfo {author} {\bibfnamefont
  {M.}~\bibnamefont {{Kotschenreuther}}}, \bibinfo {author} {\bibfnamefont
  {A.~H.}\ \bibnamefont {{Kritz}}}, \bibinfo {author} {\bibfnamefont {L.~L.}\
  \bibnamefont {{Lao}}}, \bibinfo {author} {\bibfnamefont {J.}~\bibnamefont
  {{Mandrekas}}}, \bibinfo {author} {\bibfnamefont {W.~M.}\ \bibnamefont
  {{Nevins}}}, \bibinfo {author} {\bibfnamefont {S.~E.}\ \bibnamefont
  {{Parker}}}, \bibinfo {author} {\bibfnamefont {A.~J.}\ \bibnamefont
  {{Redd}}}, \bibinfo {author} {\bibfnamefont {D.~E.}\ \bibnamefont
  {{Shumaker}}}, \bibinfo {author} {\bibfnamefont {R.}~\bibnamefont
  {{Sydora}}}, \ and\ \bibinfo {author} {\bibfnamefont {J.}~\bibnamefont
  {{Weiland}}},\ }\href {\doibase 10.1063/1.873896} {\bibfield  {journal}
  {\bibinfo  {journal} {Phys. Plasmas}\ }\textbf {\bibinfo {volume} {7}},\
  \bibinfo {pages} {969} (\bibinfo {year} {2000})}\BibitemShut {NoStop}%
\bibitem [{\citenamefont {{Rogers}}\ \emph {et~al.}(2000)\citenamefont
  {{Rogers}}, \citenamefont {{Dorland}},\ and\ \citenamefont
  {{Kotschenreuther}}}]{rogers_prl_2000}%
  \BibitemOpen
  \bibfield  {author} {\bibinfo {author} {\bibfnamefont {B.~N.}\ \bibnamefont
  {{Rogers}}}, \bibinfo {author} {\bibfnamefont {W.}~\bibnamefont {{Dorland}}},
  \ and\ \bibinfo {author} {\bibfnamefont {M.}~\bibnamefont
  {{Kotschenreuther}}},\ }\href {\doibase 10.1103/PhysRevLett.85.5336}
  {\bibfield  {journal} {\bibinfo  {journal} {Phys. Rev. Lett.}\ }\textbf
  {\bibinfo {volume} {85}},\ \bibinfo {pages} {5336} (\bibinfo {year}
  {2000})}\BibitemShut {NoStop}%
\bibitem [{\citenamefont {{Mikkelsen}}\ and\ \citenamefont
  {{Dorland}}(2008)}]{mikkelsen_prl_2008}%
  \BibitemOpen
  \bibfield  {author} {\bibinfo {author} {\bibfnamefont {D.~R.}\ \bibnamefont
  {{Mikkelsen}}}\ and\ \bibinfo {author} {\bibfnamefont {W.}~\bibnamefont
  {{Dorland}}},\ }\href {\doibase 10.1103/PhysRevLett.101.135003} {\bibfield
  {journal} {\bibinfo  {journal} {Phys. Rev. Lett.}\ }\textbf {\bibinfo
  {volume} {101}},\ \bibinfo {eid} {135003} (\bibinfo {year}
  {2008})}\BibitemShut {NoStop}%
\bibitem [{\citenamefont {Newton}\ \emph {et~al.}(2010)\citenamefont {Newton},
  \citenamefont {Cowley},\ and\ \citenamefont {Loureiro}}]{newton_ppcf_2010}%
  \BibitemOpen
  \bibfield  {author} {\bibinfo {author} {\bibfnamefont {S.~L.}\ \bibnamefont
  {Newton}}, \bibinfo {author} {\bibfnamefont {S.~C.}\ \bibnamefont {Cowley}},
  \ and\ \bibinfo {author} {\bibfnamefont {N.~F.}\ \bibnamefont {Loureiro}},\
  }\href@noop {} {\bibfield  {journal} {\bibinfo  {journal} {Plasma Phys.
  Control. Fusion}\ }\textbf {\bibinfo {volume} {52}},\ \bibinfo {pages}
  {125001} (\bibinfo {year} {2010})}\BibitemShut {NoStop}%
\bibitem [{\citenamefont {Schekochihin}\ \emph {et~al.}(2012)\citenamefont
  {Schekochihin}, \citenamefont {Highcock},\ and\ \citenamefont
  {Cowley}}]{schekochihin_ppcf_2012}%
  \BibitemOpen
  \bibfield  {author} {\bibinfo {author} {\bibfnamefont {A.~A.}\ \bibnamefont
  {Schekochihin}}, \bibinfo {author} {\bibfnamefont {E.~G.}\ \bibnamefont
  {Highcock}}, \ and\ \bibinfo {author} {\bibfnamefont {S.~C.}\ \bibnamefont
  {Cowley}},\ }\href@noop {} {\bibfield  {journal} {\bibinfo  {journal} {Plasma
  Phys. Control. Fusion}\ }\textbf {\bibinfo {volume} {54}},\ \bibinfo {pages}
  {055011} (\bibinfo {year} {2012})}\BibitemShut {NoStop}%
\bibitem [{\citenamefont {Barnes}\ \emph
  {et~al.}(2011{\natexlab{b}})\citenamefont {Barnes}, \citenamefont {Parra},
  \citenamefont {Highcock}, \citenamefont {Schekochihin}, \citenamefont
  {Cowley},\ and\ \citenamefont {Roach}}]{barnes_prl_2011}%
  \BibitemOpen
  \bibfield  {author} {\bibinfo {author} {\bibfnamefont {M.}~\bibnamefont
  {Barnes}}, \bibinfo {author} {\bibfnamefont {F.~I.}\ \bibnamefont {Parra}},
  \bibinfo {author} {\bibfnamefont {E.~G.}\ \bibnamefont {Highcock}}, \bibinfo
  {author} {\bibfnamefont {A.~A.}\ \bibnamefont {Schekochihin}}, \bibinfo
  {author} {\bibfnamefont {S.~C.}\ \bibnamefont {Cowley}}, \ and\ \bibinfo
  {author} {\bibfnamefont {C.~M.}\ \bibnamefont {Roach}},\ }\href@noop {}
  {\bibfield  {journal} {\bibinfo  {journal} {Phys. Rev. Lett.}\ }\textbf
  {\bibinfo {volume} {106}},\ \bibinfo {pages} {175004} (\bibinfo {year}
  {2011}{\natexlab{b}})}\BibitemShut {NoStop}%
\bibitem [{\citenamefont {Highcock}\ \emph {et~al.}(2010)\citenamefont
  {Highcock}, \citenamefont {Barnes}, \citenamefont {Schekochihin},
  \citenamefont {Parra}, \citenamefont {Roach},\ and\ \citenamefont
  {Cowley}}]{highcock_prl_2010}%
  \BibitemOpen
  \bibfield  {author} {\bibinfo {author} {\bibfnamefont {E.~G.}\ \bibnamefont
  {Highcock}}, \bibinfo {author} {\bibfnamefont {M.}~\bibnamefont {Barnes}},
  \bibinfo {author} {\bibfnamefont {A.~A.}\ \bibnamefont {Schekochihin}},
  \bibinfo {author} {\bibfnamefont {F.~I.}\ \bibnamefont {Parra}}, \bibinfo
  {author} {\bibfnamefont {C.~M.}\ \bibnamefont {Roach}}, \ and\ \bibinfo
  {author} {\bibfnamefont {S.~C.}\ \bibnamefont {Cowley}},\ }\href@noop {}
  {\bibfield  {journal} {\bibinfo  {journal} {Phys. Rev. Lett.}\ }\textbf
  {\bibinfo {volume} {105}},\ \bibinfo {pages} {215003} (\bibinfo {year}
  {2010})}\BibitemShut {NoStop}%
\bibitem [{\citenamefont {Highcock}\ \emph {et~al.}(2011)\citenamefont
  {Highcock}, \citenamefont {Barnes}, \citenamefont {Parra}, \citenamefont
  {Schekochihin}, \citenamefont {Roach},\ and\ \citenamefont
  {Cowley}}]{highcock_pop_2011}%
  \BibitemOpen
  \bibfield  {author} {\bibinfo {author} {\bibfnamefont {E.~G.}\ \bibnamefont
  {Highcock}}, \bibinfo {author} {\bibfnamefont {M.}~\bibnamefont {Barnes}},
  \bibinfo {author} {\bibfnamefont {F.~I.}\ \bibnamefont {Parra}}, \bibinfo
  {author} {\bibfnamefont {A.~A.}\ \bibnamefont {Schekochihin}}, \bibinfo
  {author} {\bibfnamefont {C.~M.}\ \bibnamefont {Roach}}, \ and\ \bibinfo
  {author} {\bibfnamefont {S.~C.}\ \bibnamefont {Cowley}},\ }\href@noop {}
  {\bibfield  {journal} {\bibinfo  {journal} {Phys. Plasmas}\ }\textbf
  {\bibinfo {volume} {18}},\ \bibinfo {pages} {102304} (\bibinfo {year}
  {2011})}\BibitemShut {NoStop}%
\bibitem [{\citenamefont {Ghim}\ \emph {et~al.}(2012)\citenamefont {Ghim},
  \citenamefont {Field}, \citenamefont {Duani}, \citenamefont {Zoletnik},
  \citenamefont {Bardoczi}, \citenamefont {Schekochihin},\ and\ \citenamefont
  {{the MAST Team}}}]{ghim_ppcf_2012}%
  \BibitemOpen
  \bibfield  {author} {\bibinfo {author} {\bibfnamefont {Y.-c.}\ \bibnamefont
  {Ghim}}, \bibinfo {author} {\bibfnamefont {A.~R.}\ \bibnamefont {Field}},
  \bibinfo {author} {\bibfnamefont {D.}~\bibnamefont {Duani}}, \bibinfo
  {author} {\bibfnamefont {S.}~\bibnamefont {Zoletnik}}, \bibinfo {author}
  {\bibfnamefont {L.}~\bibnamefont {Bardoczi}}, \bibinfo {author}
  {\bibfnamefont {A.~A.}\ \bibnamefont {Schekochihin}}, \ and\ \bibinfo
  {author} {\bibnamefont {{the MAST Team}}},\ }\href@noop {} {\bibfield
  {journal} {\bibinfo  {journal} {Plasma Phys. Control. Fusion}\ }\textbf
  {\bibinfo {volume} {54}},\ \bibinfo {pages} {095012} (\bibinfo {year}
  {2012})}\BibitemShut {NoStop}%
\bibitem [{\citenamefont {Field}\ \emph {et~al.}(2009)\citenamefont {Field},
  \citenamefont {McCone}, \citenamefont {Conway}, \citenamefont {Dunstan},
  \citenamefont {Newton},\ and\ \citenamefont {Wisse}}]{field_ppcf_2009}%
  \BibitemOpen
  \bibfield  {author} {\bibinfo {author} {\bibfnamefont {A.~R.}\ \bibnamefont
  {Field}}, \bibinfo {author} {\bibfnamefont {J.}~\bibnamefont {McCone}},
  \bibinfo {author} {\bibfnamefont {N.~J.}\ \bibnamefont {Conway}}, \bibinfo
  {author} {\bibfnamefont {M.}~\bibnamefont {Dunstan}}, \bibinfo {author}
  {\bibfnamefont {S.}~\bibnamefont {Newton}}, \ and\ \bibinfo {author}
  {\bibfnamefont {M.}~\bibnamefont {Wisse}},\ }\href@noop {} {\bibfield
  {journal} {\bibinfo  {journal} {Plasma Phys. Control. Fusion}\ }\textbf
  {\bibinfo {volume} {51}},\ \bibinfo {pages} {105002} (\bibinfo {year}
  {2009})}\BibitemShut {NoStop}%
\bibitem [{\citenamefont {Hinton}\ and\ \citenamefont
  {Wong}(1985)}]{hinton_pf_1985}%
  \BibitemOpen
  \bibfield  {author} {\bibinfo {author} {\bibfnamefont {F.~L.}\ \bibnamefont
  {Hinton}}\ and\ \bibinfo {author} {\bibfnamefont {S.~K.}\ \bibnamefont
  {Wong}},\ }\href@noop {} {\bibfield  {journal} {\bibinfo  {journal} {Phys.
  Fluids}\ }\textbf {\bibinfo {volume} {28}},\ \bibinfo {pages} {3082}
  (\bibinfo {year} {1985})}\BibitemShut {NoStop}%
\bibitem [{\citenamefont {Cowley}\ and\ \citenamefont
  {Bishop}(1986)}]{cowley_clr_1986}%
  \BibitemOpen
  \bibfield  {author} {\bibinfo {author} {\bibfnamefont {S.~C.}\ \bibnamefont
  {Cowley}}\ and\ \bibinfo {author} {\bibfnamefont {C.~M.}\ \bibnamefont
  {Bishop}},\ }\href@noop {} {\bibfield  {journal} {\bibinfo  {journal} {Culham
  Laboratory Report CLM-M}\ ,\ \bibinfo {pages} {109}} (\bibinfo {year}
  {1986})}\BibitemShut {NoStop}%
\bibitem [{Note2()}]{Note2}%
  \BibitemOpen
  \bibinfo {note} {Any mean poloidal flow exceeding the diamagnetic velocity
  would be damped by collisions \cite
  {connor_ppcf_1987,catto_pf_1987}.}\BibitemShut {Stop}%
\bibitem [{\citenamefont {Dorland}\ \emph {et~al.}(1994)\citenamefont
  {Dorland}, \citenamefont {Kotschenreuther}, \citenamefont {Beer},
  \citenamefont {Hammett}, \citenamefont {Waltz}, \citenamefont {Dominguez},
  \citenamefont {Valanju}, \citenamefont {{Miner~Jr.}}, \citenamefont {Dong},
  \citenamefont {Horton}, \citenamefont {Waelbroeck}, \citenamefont {Tajima},\
  and\ \citenamefont {LeBrun}}]{dorland_ppcnfr_1994}%
  \BibitemOpen
  \bibfield  {author} {\bibinfo {author} {\bibfnamefont {W.}~\bibnamefont
  {Dorland}}, \bibinfo {author} {\bibfnamefont {M.}~\bibnamefont
  {Kotschenreuther}}, \bibinfo {author} {\bibfnamefont {M.~A.}\ \bibnamefont
  {Beer}}, \bibinfo {author} {\bibfnamefont {G.~W.}\ \bibnamefont {Hammett}},
  \bibinfo {author} {\bibfnamefont {R.~E.}\ \bibnamefont {Waltz}}, \bibinfo
  {author} {\bibfnamefont {R.~R.}\ \bibnamefont {Dominguez}}, \bibinfo {author}
  {\bibfnamefont {P.~M.}\ \bibnamefont {Valanju}}, \bibinfo {author}
  {\bibfnamefont {W.~H.}\ \bibnamefont {{Miner~Jr.}}}, \bibinfo {author}
  {\bibfnamefont {J.~Q.}\ \bibnamefont {Dong}}, \bibinfo {author}
  {\bibfnamefont {W.}~\bibnamefont {Horton}}, \bibinfo {author} {\bibfnamefont
  {F.~L.}\ \bibnamefont {Waelbroeck}}, \bibinfo {author} {\bibfnamefont
  {T.}~\bibnamefont {Tajima}}, \ and\ \bibinfo {author} {\bibfnamefont {M.~J.}\
  \bibnamefont {LeBrun}},\ }\href@noop {} {\bibfield  {journal} {\bibinfo
  {journal} {Plasma Phys.\ Controlled Nucl.\ Fusion Res.}\ }\textbf {\bibinfo
  {volume} {3}},\ \bibinfo {pages} {463} (\bibinfo {year} {1994})}\BibitemShut
  {NoStop}%
\bibitem [{\citenamefont {Waltz}\ \emph {et~al.}(1994)\citenamefont {Waltz},
  \citenamefont {Kerbel},\ and\ \citenamefont {Milovich}}]{waltz_pop_1994}%
  \BibitemOpen
  \bibfield  {author} {\bibinfo {author} {\bibfnamefont {R.~E.}\ \bibnamefont
  {Waltz}}, \bibinfo {author} {\bibfnamefont {G.~D.}\ \bibnamefont {Kerbel}}, \
  and\ \bibinfo {author} {\bibfnamefont {J.}~\bibnamefont {Milovich}},\
  }\href@noop {} {\bibfield  {journal} {\bibinfo  {journal} {Phys. Plasmas}\
  }\textbf {\bibinfo {volume} {1}},\ \bibinfo {pages} {2229} (\bibinfo {year}
  {1994})}\BibitemShut {NoStop}%
\bibitem [{\citenamefont {Dimits}\ \emph {et~al.}(2001)\citenamefont {Dimits},
  \citenamefont {Cohen}, \citenamefont {Nevins},\ and\ \citenamefont
  {Shumaker}}]{dimits_nf_2001}%
  \BibitemOpen
  \bibfield  {author} {\bibinfo {author} {\bibfnamefont {A.~M.}\ \bibnamefont
  {Dimits}}, \bibinfo {author} {\bibfnamefont {B.~I.}\ \bibnamefont {Cohen}},
  \bibinfo {author} {\bibfnamefont {W.~M.}\ \bibnamefont {Nevins}}, \ and\
  \bibinfo {author} {\bibfnamefont {D.~E.}\ \bibnamefont {Shumaker}},\
  }\href@noop {} {\bibfield  {journal} {\bibinfo  {journal} {Nucl. Fusion}\
  }\textbf {\bibinfo {volume} {41}},\ \bibinfo {pages} {1725} (\bibinfo {year}
  {2001})}\BibitemShut {NoStop}%
\bibitem [{\citenamefont {Kinsey}\ \emph {et~al.}(2005)\citenamefont {Kinsey},
  \citenamefont {Waltz},\ and\ \citenamefont {Candy}}]{kinsey_pop_2005}%
  \BibitemOpen
  \bibfield  {author} {\bibinfo {author} {\bibfnamefont {J.~E.}\ \bibnamefont
  {Kinsey}}, \bibinfo {author} {\bibfnamefont {R.~E.}\ \bibnamefont {Waltz}}, \
  and\ \bibinfo {author} {\bibfnamefont {J.}~\bibnamefont {Candy}},\
  }\href@noop {} {\bibfield  {journal} {\bibinfo  {journal} {Phys. Plasmas}\
  }\textbf {\bibinfo {volume} {12}},\ \bibinfo {pages} {062302} (\bibinfo
  {year} {2005})}\BibitemShut {NoStop}%
\bibitem [{\citenamefont {Camenen}\ \emph {et~al.}(2009)\citenamefont
  {Camenen}, \citenamefont {Peeters}, \citenamefont {Angioni}, \citenamefont
  {Casson}, \citenamefont {Hornsby}, \citenamefont {Snodin},\ and\
  \citenamefont {Strintzi}}]{camenen_pop_2009}%
  \BibitemOpen
  \bibfield  {author} {\bibinfo {author} {\bibfnamefont {Y.}~\bibnamefont
  {Camenen}}, \bibinfo {author} {\bibfnamefont {A.~G.}\ \bibnamefont
  {Peeters}}, \bibinfo {author} {\bibfnamefont {C.}~\bibnamefont {Angioni}},
  \bibinfo {author} {\bibfnamefont {F.~J.}\ \bibnamefont {Casson}}, \bibinfo
  {author} {\bibfnamefont {W.~A.}\ \bibnamefont {Hornsby}}, \bibinfo {author}
  {\bibfnamefont {A.~P.}\ \bibnamefont {Snodin}}, \ and\ \bibinfo {author}
  {\bibfnamefont {D.}~\bibnamefont {Strintzi}},\ }\href@noop {} {\bibfield
  {journal} {\bibinfo  {journal} {Phys. Plasmas}\ }\textbf {\bibinfo {volume}
  {16}},\ \bibinfo {pages} {012503} (\bibinfo {year} {2009})}\BibitemShut
  {NoStop}%
\bibitem [{\citenamefont {Roach}\ \emph {et~al.}(2009)\citenamefont {Roach},
  \citenamefont {Abel}, \citenamefont {Akers}, \citenamefont {Arter},
  \citenamefont {Barnes}, \citenamefont {Camenen}, \citenamefont {Casson},
  \citenamefont {Colyer}, \citenamefont {Connor}, \citenamefont {Cowley},
  \citenamefont {Dickinson}, \citenamefont {Dorland}, \citenamefont {Field},
  \citenamefont {Guttenfelder}, \citenamefont {Hammett}, \citenamefont
  {Hastie}, \citenamefont {Highcock}, \citenamefont {Loureiro}, \citenamefont
  {Peeters}, \citenamefont {Reshko}, \citenamefont {Saarelma}, \citenamefont
  {Schekochihin}, \citenamefont {Valovic},\ and\ \citenamefont
  {Wilson}}]{roach_ppcf_2009}%
  \BibitemOpen
  \bibfield  {author} {\bibinfo {author} {\bibfnamefont {C.~M.}\ \bibnamefont
  {Roach}}, \bibinfo {author} {\bibfnamefont {I.~G.}\ \bibnamefont {Abel}},
  \bibinfo {author} {\bibfnamefont {R.~J.}\ \bibnamefont {Akers}}, \bibinfo
  {author} {\bibfnamefont {W.}~\bibnamefont {Arter}}, \bibinfo {author}
  {\bibfnamefont {M.}~\bibnamefont {Barnes}}, \bibinfo {author} {\bibfnamefont
  {Y.}~\bibnamefont {Camenen}}, \bibinfo {author} {\bibfnamefont {F.~J.}\
  \bibnamefont {Casson}}, \bibinfo {author} {\bibfnamefont {G.}~\bibnamefont
  {Colyer}}, \bibinfo {author} {\bibfnamefont {J.~W.}\ \bibnamefont {Connor}},
  \bibinfo {author} {\bibfnamefont {S.~C.}\ \bibnamefont {Cowley}}, \bibinfo
  {author} {\bibfnamefont {D.}~\bibnamefont {Dickinson}}, \bibinfo {author}
  {\bibfnamefont {W.}~\bibnamefont {Dorland}}, \bibinfo {author} {\bibfnamefont
  {A.~R.}\ \bibnamefont {Field}}, \bibinfo {author} {\bibfnamefont
  {W.}~\bibnamefont {Guttenfelder}}, \bibinfo {author} {\bibfnamefont {G.~W.}\
  \bibnamefont {Hammett}}, \bibinfo {author} {\bibfnamefont {R.~J.}\
  \bibnamefont {Hastie}}, \bibinfo {author} {\bibfnamefont {E.}~\bibnamefont
  {Highcock}}, \bibinfo {author} {\bibfnamefont {N.~F.}\ \bibnamefont
  {Loureiro}}, \bibinfo {author} {\bibfnamefont {A.~G.}\ \bibnamefont
  {Peeters}}, \bibinfo {author} {\bibfnamefont {M.}~\bibnamefont {Reshko}},
  \bibinfo {author} {\bibfnamefont {S.}~\bibnamefont {Saarelma}}, \bibinfo
  {author} {\bibfnamefont {A.~A.}\ \bibnamefont {Schekochihin}}, \bibinfo
  {author} {\bibfnamefont {M.}~\bibnamefont {Valovic}}, \ and\ \bibinfo
  {author} {\bibfnamefont {H.~R.}\ \bibnamefont {Wilson}},\ }\href@noop {}
  {\bibfield  {journal} {\bibinfo  {journal} {Plasma Phys. Control. Fusion}\
  }\textbf {\bibinfo {volume} {51}},\ \bibinfo {pages} {124020} (\bibinfo
  {year} {2009})}\BibitemShut {NoStop}%
\bibitem [{\citenamefont {Casson}\ \emph {et~al.}(2009)\citenamefont {Casson},
  \citenamefont {Peeters}, \citenamefont {Camenen}, \citenamefont {Hornsby},
  \citenamefont {Snodin}, \citenamefont {Strintzi},\ and\ \citenamefont
  {Szepesi}}]{casson_pop_2009}%
  \BibitemOpen
  \bibfield  {author} {\bibinfo {author} {\bibfnamefont {F.~J.}\ \bibnamefont
  {Casson}}, \bibinfo {author} {\bibfnamefont {A.~G.}\ \bibnamefont {Peeters}},
  \bibinfo {author} {\bibfnamefont {Y.}~\bibnamefont {Camenen}}, \bibinfo
  {author} {\bibfnamefont {W.~A.}\ \bibnamefont {Hornsby}}, \bibinfo {author}
  {\bibfnamefont {A.~P.}\ \bibnamefont {Snodin}}, \bibinfo {author}
  {\bibfnamefont {D.}~\bibnamefont {Strintzi}}, \ and\ \bibinfo {author}
  {\bibfnamefont {G.}~\bibnamefont {Szepesi}},\ }\href@noop {} {\ \textbf
  {\bibinfo {volume} {16}},\ \bibinfo {pages} {092303} (\bibinfo {year}
  {2009})}\BibitemShut {NoStop}%
\bibitem [{\citenamefont {Burrell}(1997)}]{burrell_pop_1997}%
  \BibitemOpen
  \bibfield  {author} {\bibinfo {author} {\bibfnamefont {K.~H.}\ \bibnamefont
  {Burrell}},\ }\href@noop {} {\bibfield  {journal} {\bibinfo  {journal} {Phys.
  Plasmas}\ }\textbf {\bibinfo {volume} {4}},\ \bibinfo {pages} {1499}
  (\bibinfo {year} {1997})}\BibitemShut {NoStop}%
\bibitem [{\citenamefont {Burrell}(1999)}]{burrell_pop_1999}%
  \BibitemOpen
  \bibfield  {author} {\bibinfo {author} {\bibfnamefont {K.~H.}\ \bibnamefont
  {Burrell}},\ }\href@noop {} {\bibfield  {journal} {\bibinfo  {journal} {Phys.
  Plasmas}\ }\textbf {\bibinfo {volume} {6}},\ \bibinfo {pages} {4418}
  (\bibinfo {year} {1999})}\BibitemShut {NoStop}%
\bibitem [{\citenamefont {{Schaffner}}\ \emph {et~al.}(2012)\citenamefont
  {{Schaffner}}, \citenamefont {{Carter}}, \citenamefont {{Rossi}},
  \citenamefont {{Guice}}, \citenamefont {{Maggs}}, \citenamefont {{Vincena}},\
  and\ \citenamefont {{Friedman}}}]{schaffner_prl_2012}%
  \BibitemOpen
  \bibfield  {author} {\bibinfo {author} {\bibfnamefont {D.~A.}\ \bibnamefont
  {{Schaffner}}}, \bibinfo {author} {\bibfnamefont {T.~A.}\ \bibnamefont
  {{Carter}}}, \bibinfo {author} {\bibfnamefont {G.~D.}\ \bibnamefont
  {{Rossi}}}, \bibinfo {author} {\bibfnamefont {D.~S.}\ \bibnamefont
  {{Guice}}}, \bibinfo {author} {\bibfnamefont {J.~E.}\ \bibnamefont
  {{Maggs}}}, \bibinfo {author} {\bibfnamefont {S.}~\bibnamefont {{Vincena}}},
  \ and\ \bibinfo {author} {\bibfnamefont {B.}~\bibnamefont {{Friedman}}},\
  }\href {\doibase 10.1103/PhysRevLett.109.135002} {\bibfield  {journal}
  {\bibinfo  {journal} {Phys. Rev. Lett.}\ }\textbf {\bibinfo {volume} {109}},\
  \bibinfo {eid} {135002} (\bibinfo {year} {2012})}\BibitemShut {NoStop}%
\bibitem [{\citenamefont {Catto}\ \emph {et~al.}(1973)\citenamefont {Catto},
  \citenamefont {Rosenbluth},\ and\ \citenamefont {Liu}}]{catto_pf_1973}%
  \BibitemOpen
  \bibfield  {author} {\bibinfo {author} {\bibfnamefont {P.~J.}\ \bibnamefont
  {Catto}}, \bibinfo {author} {\bibfnamefont {M.~N.}\ \bibnamefont
  {Rosenbluth}}, \ and\ \bibinfo {author} {\bibfnamefont {C.~S.}\ \bibnamefont
  {Liu}},\ }\href@noop {} {\bibfield  {journal} {\bibinfo  {journal} {Phys.
  Fluids}\ }\textbf {\bibinfo {volume} {16}},\ \bibinfo {pages} {1719}
  (\bibinfo {year} {1973})}\BibitemShut {NoStop}%
\bibitem [{\citenamefont {Field}\ \emph {et~al.}(2012)\citenamefont {Field},
  \citenamefont {Dunai}, \citenamefont {Gaffka}, \citenamefont {Ghim},
  \citenamefont {Kiss}, \citenamefont {Meszaros}, \citenamefont {Krizsanoczi},
  \citenamefont {Shibaev},\ and\ \citenamefont {Zoletnik}}]{field_rsi_2012}%
  \BibitemOpen
  \bibfield  {author} {\bibinfo {author} {\bibfnamefont {A.~R.}\ \bibnamefont
  {Field}}, \bibinfo {author} {\bibfnamefont {D.}~\bibnamefont {Dunai}},
  \bibinfo {author} {\bibfnamefont {R.}~\bibnamefont {Gaffka}}, \bibinfo
  {author} {\bibfnamefont {Y.-c.}\ \bibnamefont {Ghim}}, \bibinfo {author}
  {\bibfnamefont {I.}~\bibnamefont {Kiss}}, \bibinfo {author} {\bibfnamefont
  {B.}~\bibnamefont {Meszaros}}, \bibinfo {author} {\bibfnamefont
  {T.}~\bibnamefont {Krizsanoczi}}, \bibinfo {author} {\bibfnamefont
  {S.}~\bibnamefont {Shibaev}}, \ and\ \bibinfo {author} {\bibfnamefont
  {S.}~\bibnamefont {Zoletnik}},\ }\href@noop {} {\bibfield  {journal}
  {\bibinfo  {journal} {Rev. Sci. Instrum.}\ }\textbf {\bibinfo {volume}
  {83}},\ \bibinfo {pages} {013508} (\bibinfo {year} {2012})}\BibitemShut
  {NoStop}%
\bibitem [{\citenamefont {Scannell}\ \emph {et~al.}(2010)\citenamefont
  {Scannell}, \citenamefont {Walsh}, \citenamefont {Dunstan}, \citenamefont
  {Figueiredo}, \citenamefont {Naylor}, \citenamefont {O'Gorman}, \citenamefont
  {Shibaev}, \citenamefont {Gibson},\ and\ \citenamefont
  {Wilson}}]{scannell_rsi_2010}%
  \BibitemOpen
  \bibfield  {author} {\bibinfo {author} {\bibfnamefont {R.}~\bibnamefont
  {Scannell}}, \bibinfo {author} {\bibfnamefont {M.~J.}\ \bibnamefont {Walsh}},
  \bibinfo {author} {\bibfnamefont {M.~R.}\ \bibnamefont {Dunstan}}, \bibinfo
  {author} {\bibfnamefont {J.}~\bibnamefont {Figueiredo}}, \bibinfo {author}
  {\bibfnamefont {G.}~\bibnamefont {Naylor}}, \bibinfo {author} {\bibfnamefont
  {T.}~\bibnamefont {O'Gorman}}, \bibinfo {author} {\bibfnamefont
  {S.}~\bibnamefont {Shibaev}}, \bibinfo {author} {\bibfnamefont {K.~J.}\
  \bibnamefont {Gibson}}, \ and\ \bibinfo {author} {\bibfnamefont
  {H.}~\bibnamefont {Wilson}},\ }\href@noop {} {\bibfield  {journal} {\bibinfo
  {journal} {Rev. Sci. Instrum.}\ }\textbf {\bibinfo {volume} {81}},\ \bibinfo
  {pages} {10D520} (\bibinfo {year} {2010})}\BibitemShut {NoStop}%
\bibitem [{\citenamefont {Conway}\ \emph {et~al.}(2006)\citenamefont {Conway},
  \citenamefont {Carolan}, \citenamefont {McCone}, \citenamefont {Walsh},\ and\
  \citenamefont {Wisse}}]{conway_rsi_2006}%
  \BibitemOpen
  \bibfield  {author} {\bibinfo {author} {\bibfnamefont {N.~J.}\ \bibnamefont
  {Conway}}, \bibinfo {author} {\bibfnamefont {P.~G.}\ \bibnamefont {Carolan}},
  \bibinfo {author} {\bibfnamefont {J.}~\bibnamefont {McCone}}, \bibinfo
  {author} {\bibfnamefont {M.~J.}\ \bibnamefont {Walsh}}, \ and\ \bibinfo
  {author} {\bibfnamefont {M.}~\bibnamefont {Wisse}},\ }\href@noop {}
  {\bibfield  {journal} {\bibinfo  {journal} {Rev. Sci. Instrum.}\ }\textbf
  {\bibinfo {volume} {77}},\ \bibinfo {pages} {10F131} (\bibinfo {year}
  {2006})}\BibitemShut {NoStop}%
\bibitem [{\citenamefont {Kim}\ \emph {et~al.}(1990)\citenamefont {Kim},
  \citenamefont {Diamond}, \citenamefont {Biglari},\ and\ \citenamefont
  {Terry}}]{kim_pfb_1990}%
  \BibitemOpen
  \bibfield  {author} {\bibinfo {author} {\bibfnamefont {Y.~B.}\ \bibnamefont
  {Kim}}, \bibinfo {author} {\bibfnamefont {P.~H.}\ \bibnamefont {Diamond}},
  \bibinfo {author} {\bibfnamefont {H.}~\bibnamefont {Biglari}}, \ and\
  \bibinfo {author} {\bibfnamefont {P.~W.}\ \bibnamefont {Terry}},\ }\href@noop
  {} {\bibfield  {journal} {\bibinfo  {journal} {Phys. Fluids B}\ }\textbf
  {\bibinfo {volume} {2}},\ \bibinfo {pages} {2143} (\bibinfo {year}
  {1990})}\BibitemShut {NoStop}%
\bibitem [{\citenamefont {De~Bock}\ \emph {et~al.}(2008)\citenamefont
  {De~Bock}, \citenamefont {Conway}, \citenamefont {Walsh}, \citenamefont
  {Carolan},\ and\ \citenamefont {Hawkes}}]{debock_rsi_2008}%
  \BibitemOpen
  \bibfield  {author} {\bibinfo {author} {\bibfnamefont {M.~F.~M.}\
  \bibnamefont {De~Bock}}, \bibinfo {author} {\bibfnamefont {N.~J.}\
  \bibnamefont {Conway}}, \bibinfo {author} {\bibfnamefont {M.~J.}\
  \bibnamefont {Walsh}}, \bibinfo {author} {\bibfnamefont {P.~G.}\ \bibnamefont
  {Carolan}}, \ and\ \bibinfo {author} {\bibfnamefont {N.~C.}\ \bibnamefont
  {Hawkes}},\ }\href@noop {} {\bibfield  {journal} {\bibinfo  {journal} {Rev.
  Sci. Instrum.}\ }\textbf {\bibinfo {volume} {79}},\ \bibinfo {pages} {10F524}
  (\bibinfo {year} {2008})}\BibitemShut {NoStop}%
\bibitem [{\citenamefont {Lao}\ \emph {et~al.}(1985)\citenamefont {Lao},
  \citenamefont {St~John}, \citenamefont {Stambaugh}, \citenamefont {Kellman},\
  and\ \citenamefont {Pfeiffer}}]{lao_nf_1985}%
  \BibitemOpen
  \bibfield  {author} {\bibinfo {author} {\bibfnamefont {L.~L.}\ \bibnamefont
  {Lao}}, \bibinfo {author} {\bibfnamefont {H.}~\bibnamefont {St~John}},
  \bibinfo {author} {\bibfnamefont {R.~D.}\ \bibnamefont {Stambaugh}}, \bibinfo
  {author} {\bibfnamefont {A.~G.}\ \bibnamefont {Kellman}}, \ and\ \bibinfo
  {author} {\bibfnamefont {W.}~\bibnamefont {Pfeiffer}},\ }\href@noop {}
  {\bibfield  {journal} {\bibinfo  {journal} {Nucl. Fusion}\ }\textbf {\bibinfo
  {volume} {25}},\ \bibinfo {pages} {1611} (\bibinfo {year}
  {1985})}\BibitemShut {NoStop}%
\bibitem [{\citenamefont {{Beer}}\ \emph {et~al.}(1997)\citenamefont {{Beer}},
  \citenamefont {{Hammett}}, \citenamefont {{Rewoldt}}, \citenamefont
  {{Synakowski}}, \citenamefont {{Zarnstorff}},\ and\ \citenamefont
  {{Dorland}}}]{beer_pop_1997}%
  \BibitemOpen
  \bibfield  {author} {\bibinfo {author} {\bibfnamefont {M.~A.}\ \bibnamefont
  {{Beer}}}, \bibinfo {author} {\bibfnamefont {G.~W.}\ \bibnamefont
  {{Hammett}}}, \bibinfo {author} {\bibfnamefont {G.}~\bibnamefont
  {{Rewoldt}}}, \bibinfo {author} {\bibfnamefont {E.~J.}\ \bibnamefont
  {{Synakowski}}}, \bibinfo {author} {\bibfnamefont {M.~C.}\ \bibnamefont
  {{Zarnstorff}}}, \ and\ \bibinfo {author} {\bibfnamefont {W.}~\bibnamefont
  {{Dorland}}},\ }\href {\doibase 10.1063/1.872279} {\bibfield  {journal}
  {\bibinfo  {journal} {Phys. Plasmas}\ }\textbf {\bibinfo {volume} {4}},\
  \bibinfo {pages} {1792} (\bibinfo {year} {1997})}\BibitemShut {NoStop}%
\bibitem [{\citenamefont {{Rechester}}\ and\ \citenamefont
  {{Rosenbluth}}(1978)}]{rechester_prl_1978}%
  \BibitemOpen
  \bibfield  {author} {\bibinfo {author} {\bibfnamefont {A.~B.}\ \bibnamefont
  {{Rechester}}}\ and\ \bibinfo {author} {\bibfnamefont {M.~N.}\ \bibnamefont
  {{Rosenbluth}}},\ }\href {\doibase 10.1103/PhysRevLett.40.38} {\bibfield
  {journal} {\bibinfo  {journal} {Phys. Rev. Lett.}\ }\textbf {\bibinfo
  {volume} {40}},\ \bibinfo {pages} {38} (\bibinfo {year} {1978})}\BibitemShut
  {NoStop}%
\bibitem [{\citenamefont {Pueschel}\ and\ \citenamefont
  {Jenko}(2010)}]{pueschel_pop_2010}%
  \BibitemOpen
  \bibfield  {author} {\bibinfo {author} {\bibfnamefont {M.~J.}\ \bibnamefont
  {Pueschel}}\ and\ \bibinfo {author} {\bibfnamefont {F.}~\bibnamefont
  {Jenko}},\ }\href@noop {} {\bibfield  {journal} {\bibinfo  {journal} {Phys.
  Plasmas}\ }\textbf {\bibinfo {volume} {17}},\ \bibinfo {pages} {062307}
  (\bibinfo {year} {2010})}\BibitemShut {NoStop}%
\bibitem [{\citenamefont {Nevins}\ \emph {et~al.}(2011)\citenamefont {Nevins},
  \citenamefont {Wang},\ and\ \citenamefont {Candy}}]{nevins_prl_2011}%
  \BibitemOpen
  \bibfield  {author} {\bibinfo {author} {\bibfnamefont {W.~M.}\ \bibnamefont
  {Nevins}}, \bibinfo {author} {\bibfnamefont {E.}~\bibnamefont {Wang}}, \ and\
  \bibinfo {author} {\bibfnamefont {J.}~\bibnamefont {Candy}},\ }\href@noop {}
  {\bibfield  {journal} {\bibinfo  {journal} {Phys. Rev. Lett.}\ }\textbf
  {\bibinfo {volume} {106}},\ \bibinfo {pages} {065003} (\bibinfo {year}
  {2011})}\BibitemShut {NoStop}%
\bibitem [{\citenamefont {Hatch}\ \emph {et~al.}(2012)\citenamefont {Hatch},
  \citenamefont {Pueschel}, \citenamefont {Jenko}, \citenamefont {Nevins},
  \citenamefont {Terry},\ and\ \citenamefont {Doerk}}]{hatch_prl_2012}%
  \BibitemOpen
  \bibfield  {author} {\bibinfo {author} {\bibfnamefont {D.~R.}\ \bibnamefont
  {Hatch}}, \bibinfo {author} {\bibfnamefont {M.~J.}\ \bibnamefont {Pueschel}},
  \bibinfo {author} {\bibfnamefont {F.}~\bibnamefont {Jenko}}, \bibinfo
  {author} {\bibfnamefont {W.~M.}\ \bibnamefont {Nevins}}, \bibinfo {author}
  {\bibfnamefont {P.~W.}\ \bibnamefont {Terry}}, \ and\ \bibinfo {author}
  {\bibfnamefont {H.}~\bibnamefont {Doerk}},\ }\href@noop {} {\bibfield
  {journal} {\bibinfo  {journal} {Phys. Rev. Lett.}\ }\textbf {\bibinfo
  {volume} {108}},\ \bibinfo {pages} {235002} (\bibinfo {year}
  {2012})}\BibitemShut {NoStop}%
\bibitem [{\citenamefont {{Guttenfelder}}\ \emph {et~al.}(2012)\citenamefont
  {{Guttenfelder}}, \citenamefont {{Candy}}, \citenamefont {{Kaye}},
  \citenamefont {{Nevins}}, \citenamefont {{Wang}}, \citenamefont {{Zhang}},
  \citenamefont {{Bell}}, \citenamefont {{Crocker}}, \citenamefont {{Hammett}},
  \citenamefont {{LeBlanc}}, \citenamefont {{Mikkelsen}}, \citenamefont
  {{Ren}},\ and\ \citenamefont {{Yuh}}}]{guttenfelder_pop_2012}%
  \BibitemOpen
  \bibfield  {author} {\bibinfo {author} {\bibfnamefont {W.}~\bibnamefont
  {{Guttenfelder}}}, \bibinfo {author} {\bibfnamefont {J.}~\bibnamefont
  {{Candy}}}, \bibinfo {author} {\bibfnamefont {S.~M.}\ \bibnamefont {{Kaye}}},
  \bibinfo {author} {\bibfnamefont {W.~M.}\ \bibnamefont {{Nevins}}}, \bibinfo
  {author} {\bibfnamefont {E.}~\bibnamefont {{Wang}}}, \bibinfo {author}
  {\bibfnamefont {J.}~\bibnamefont {{Zhang}}}, \bibinfo {author} {\bibfnamefont
  {R.~E.}\ \bibnamefont {{Bell}}}, \bibinfo {author} {\bibfnamefont {N.~A.}\
  \bibnamefont {{Crocker}}}, \bibinfo {author} {\bibfnamefont {G.~W.}\
  \bibnamefont {{Hammett}}}, \bibinfo {author} {\bibfnamefont {B.~P.}\
  \bibnamefont {{LeBlanc}}}, \bibinfo {author} {\bibfnamefont {D.~R.}\
  \bibnamefont {{Mikkelsen}}}, \bibinfo {author} {\bibfnamefont
  {Y.}~\bibnamefont {{Ren}}}, \ and\ \bibinfo {author} {\bibfnamefont
  {H.}~\bibnamefont {{Yuh}}},\ }\href {\doibase 10.1063/1.3694104} {\bibfield
  {journal} {\bibinfo  {journal} {Phys. Plasmas}\ }\textbf {\bibinfo {volume}
  {19}},\ \bibinfo {pages} {056119} (\bibinfo {year} {2012})}\BibitemShut
  {NoStop}%
\bibitem [{\citenamefont {{Doerk}}\ \emph {et~al.}(2012)\citenamefont
  {{Doerk}}, \citenamefont {{Jenko}}, \citenamefont {{G{\"o}rler}},
  \citenamefont {{Told}}, \citenamefont {{Pueschel}},\ and\ \citenamefont
  {{Hatch}}}]{doerk_pop_2012}%
  \BibitemOpen
  \bibfield  {author} {\bibinfo {author} {\bibfnamefont {H.}~\bibnamefont
  {{Doerk}}}, \bibinfo {author} {\bibfnamefont {F.}~\bibnamefont {{Jenko}}},
  \bibinfo {author} {\bibfnamefont {T.}~\bibnamefont {{G{\"o}rler}}}, \bibinfo
  {author} {\bibfnamefont {D.}~\bibnamefont {{Told}}}, \bibinfo {author}
  {\bibfnamefont {M.~J.}\ \bibnamefont {{Pueschel}}}, \ and\ \bibinfo {author}
  {\bibfnamefont {D.~R.}\ \bibnamefont {{Hatch}}},\ }\href {\doibase
  10.1063/1.3694663} {\bibfield  {journal} {\bibinfo  {journal} {Phys.
  Plasmas}\ }\textbf {\bibinfo {volume} {19}},\ \bibinfo {pages} {055907}
  (\bibinfo {year} {2012})}\BibitemShut {NoStop}%
\bibitem [{\citenamefont {{Abel}}\ and\ \citenamefont
  {{Cowley}}(2012)}]{abel_njp_2012}%
  \BibitemOpen
  \bibfield  {author} {\bibinfo {author} {\bibfnamefont {I.~G.}\ \bibnamefont
  {{Abel}}}\ and\ \bibinfo {author} {\bibfnamefont {S.~C.}\ \bibnamefont
  {{Cowley}}},\ }\href@noop {} {\bibfield  {journal} {\bibinfo  {journal}
  {submitted to New J. Phys. [arXiv:1210.1417]}\ } (\bibinfo {year}
  {2012})}\BibitemShut {NoStop}%
\bibitem [{\citenamefont {{Hotelling}}(1936)}]{hotelling_bio_1936}%
  \BibitemOpen
  \bibfield  {author} {\bibinfo {author} {\bibfnamefont {H.}~\bibnamefont
  {{Hotelling}}},\ }\href@noop {} {\bibfield  {journal} {\bibinfo  {journal}
  {Biometrika}\ }\textbf {\bibinfo {volume} {28}},\ \bibinfo {pages} {321}
  (\bibinfo {year} {1936})}\BibitemShut {NoStop}%
\bibitem [{Note3()}]{Note3}%
  \BibitemOpen
  \bibinfo {note} {The broad scatter of data points in Fig.~\ref
  {fig:q_eps_tau_st_tau_sh_R_LTi}(a) suggests that the correlation between
  $q/\varepsilon $ and $\protect \mathaccentV {bar}016\gamma _E$ is weak; the
  lack of higher values of $\protect \mathaccentV {bar}016\gamma _E$ at large
  $q/\varepsilon $ is due to the fact that the flow shear is weak at earlier
  times in the discharges, when the central value of $q$ is high.}\BibitemShut
  {Stop}%
\bibitem [{Note4()}]{Note4}%
  \BibitemOpen
  \bibinfo {note} {It can be proven that the functional dependence $q(\psi )$,
  where $\psi $ is the flux-surface label, only changes on the resistive
  timescale of the mean magnetic field \cite {abel_njp_2012}.}\BibitemShut
  {Stop}%
\bibitem [{\citenamefont {{Parra}}\ \emph {et~al.}(2011)\citenamefont
  {{Parra}}, \citenamefont {{Barnes}}, \citenamefont {{Highcock}},
  \citenamefont {{Schekochihin}},\ and\ \citenamefont
  {{Cowley}}}]{parra_prl_2011}%
  \BibitemOpen
  \bibfield  {author} {\bibinfo {author} {\bibfnamefont {F.~I.}\ \bibnamefont
  {{Parra}}}, \bibinfo {author} {\bibfnamefont {M.}~\bibnamefont {{Barnes}}},
  \bibinfo {author} {\bibfnamefont {E.~G.}\ \bibnamefont {{Highcock}}},
  \bibinfo {author} {\bibfnamefont {A.~A.}\ \bibnamefont {{Schekochihin}}}, \
  and\ \bibinfo {author} {\bibfnamefont {S.~C.}\ \bibnamefont {{Cowley}}},\
  }\href {\doibase 10.1103/PhysRevLett.106.115004} {\bibfield  {journal}
  {\bibinfo  {journal} {Phys. Rev. Lett.}\ }\textbf {\bibinfo {volume} {106}},\
  \bibinfo {eid} {115004} (\bibinfo {year} {2011})}\BibitemShut {NoStop}%
\bibitem [{\citenamefont {Diamond}\ \emph {et~al.}(2005)\citenamefont
  {Diamond}, \citenamefont {Itoh}, \citenamefont {Itoh},\ and\ \citenamefont
  {Hahm}}]{diamond_ppcfreview_2005}%
  \BibitemOpen
  \bibfield  {author} {\bibinfo {author} {\bibfnamefont {P.~H.}\ \bibnamefont
  {Diamond}}, \bibinfo {author} {\bibfnamefont {S.-I.}\ \bibnamefont {Itoh}},
  \bibinfo {author} {\bibfnamefont {K.}~\bibnamefont {Itoh}}, \ and\ \bibinfo
  {author} {\bibfnamefont {T.-S.}\ \bibnamefont {Hahm}},\ }\href@noop {}
  {\bibfield  {journal} {\bibinfo  {journal} {Plasma Phys. Control. Fusion}\
  }\textbf {\bibinfo {volume} {47}},\ \bibinfo {pages} {R35} (\bibinfo {year}
  {2005})}\BibitemShut {NoStop}%
\bibitem [{\citenamefont {Hinton}\ and\ \citenamefont
  {Rosenbluth}(1999)}]{hinton_ppcf_1999}%
  \BibitemOpen
  \bibfield  {author} {\bibinfo {author} {\bibfnamefont {F.~L.}\ \bibnamefont
  {Hinton}}\ and\ \bibinfo {author} {\bibfnamefont {M.~N.}\ \bibnamefont
  {Rosenbluth}},\ }\href@noop {} {\bibfield  {journal} {\bibinfo  {journal}
  {Plasma Phys. Control. Fusion}\ }\textbf {\bibinfo {volume} {41}},\ \bibinfo
  {pages} {A653} (\bibinfo {year} {1999})}\BibitemShut {NoStop}%
\bibitem [{\citenamefont {{Lin}}\ \emph {et~al.}(1999)\citenamefont {{Lin}},
  \citenamefont {{Hahm}}, \citenamefont {{Lee}}, \citenamefont {{Tang}},\ and\
  \citenamefont {{Diamond}}}]{lin_prl_1999}%
  \BibitemOpen
  \bibfield  {author} {\bibinfo {author} {\bibfnamefont {Z.}~\bibnamefont
  {{Lin}}}, \bibinfo {author} {\bibfnamefont {T.~S.}\ \bibnamefont {{Hahm}}},
  \bibinfo {author} {\bibfnamefont {W.~W.}\ \bibnamefont {{Lee}}}, \bibinfo
  {author} {\bibfnamefont {W.~M.}\ \bibnamefont {{Tang}}}, \ and\ \bibinfo
  {author} {\bibfnamefont {P.~H.}\ \bibnamefont {{Diamond}}},\ }\href {\doibase
  10.1103/PhysRevLett.83.3645} {\bibfield  {journal} {\bibinfo  {journal}
  {Phys. Rev. Lett.}\ }\textbf {\bibinfo {volume} {83}},\ \bibinfo {pages}
  {3645} (\bibinfo {year} {1999})}\BibitemShut {NoStop}%
\bibitem [{\citenamefont {Ricci}\ \emph {et~al.}(2006)\citenamefont {Ricci},
  \citenamefont {Rogers},\ and\ \citenamefont {Dorland}}]{ricci_prl_2006}%
  \BibitemOpen
  \bibfield  {author} {\bibinfo {author} {\bibfnamefont {P.}~\bibnamefont
  {Ricci}}, \bibinfo {author} {\bibfnamefont {B.~N.}\ \bibnamefont {Rogers}}, \
  and\ \bibinfo {author} {\bibfnamefont {W.}~\bibnamefont {Dorland}},\
  }\href@noop {} {\bibfield  {journal} {\bibinfo  {journal} {Phys. Rev. Lett.}\
  }\textbf {\bibinfo {volume} {97}},\ \bibinfo {pages} {245001} (\bibinfo
  {year} {2006})}\BibitemShut {NoStop}%
\bibitem [{\citenamefont {{Xiao}}\ \emph {et~al.}(2007)\citenamefont {{Xiao}},
  \citenamefont {{Catto}},\ and\ \citenamefont {{Molvig}}}]{xiao_pop_2007}%
  \BibitemOpen
  \bibfield  {author} {\bibinfo {author} {\bibfnamefont {Y.}~\bibnamefont
  {{Xiao}}}, \bibinfo {author} {\bibfnamefont {P.~J.}\ \bibnamefont {{Catto}}},
  \ and\ \bibinfo {author} {\bibfnamefont {K.}~\bibnamefont {{Molvig}}},\
  }\href {\doibase 10.1063/1.2536297} {\bibfield  {journal} {\bibinfo
  {journal} {Phys. Plasmas}\ }\textbf {\bibinfo {volume} {14}},\ \bibinfo
  {pages} {032302} (\bibinfo {year} {2007})}\BibitemShut {NoStop}%
\bibitem [{\citenamefont {{Hawryluk}}(1980)}]{hawryluk_TRANSP_1980}%
  \BibitemOpen
  \bibfield  {author} {\bibinfo {author} {\bibfnamefont {R.}~\bibnamefont
  {{Hawryluk}}},\ }in\ \href@noop {} {\emph {\bibinfo {booktitle} {Physics of
  plasmas close to thermonuclear conditions}}},\ Vol.~\bibinfo {volume} {1},\
  \bibinfo {editor} {edited by\ \bibinfo {editor} {\bibfnamefont
  {B.}~\bibnamefont {{Coppi}}}}\ (\bibinfo {organization} {Commission of the
  European Communities, Brussels},\ \bibinfo {year} {1980})\ p.~\bibinfo
  {pages} {19}\BibitemShut {NoStop}%
\bibitem [{\citenamefont {Meyer}\ \emph {et~al.}(2009)\citenamefont {Meyer}
  \emph {et~al.}}]{meyer_nf_2009}%
  \BibitemOpen
  \bibfield  {author} {\bibinfo {author} {\bibfnamefont {H.}~\bibnamefont
  {Meyer}} \emph {et~al.},\ }\href@noop {} {\bibfield  {journal} {\bibinfo
  {journal} {Nucl. Fusion}\ }\textbf {\bibinfo {volume} {49}},\ \bibinfo
  {pages} {104017} (\bibinfo {year} {2009})}\BibitemShut {NoStop}%
\bibitem [{Note5()}]{Note5}%
  \BibitemOpen
  \bibinfo {note} {This situation is distinct from the experiments on transport
  stiffness on JET \cite {mantica_prl_2009,mantica_prl_2011,mantica_ppcf_2011}
  in that they provided vigorous extra heating power using localized
  ion-cyclotron-resonance heating (ICRH) to depart far from
  marginality.}\BibitemShut {Stop}%
\bibitem [{\citenamefont {Fern}\ \emph {et~al.}(2012)\citenamefont {Fern},
  \citenamefont {Highcock}, \citenamefont {Schekochihin},\ and\ \citenamefont
  {Beurskens}}]{fern_unpub_2012}%
  \BibitemOpen
  \bibfield  {author} {\bibinfo {author} {\bibfnamefont {R.}~\bibnamefont
  {Fern}}, \bibinfo {author} {\bibfnamefont {E.~G.}\ \bibnamefont {Highcock}},
  \bibinfo {author} {\bibfnamefont {A.~A.}\ \bibnamefont {Schekochihin}}, \
  and\ \bibinfo {author} {\bibfnamefont {M.~N.~A.}\ \bibnamefont {Beurskens}},\
  }\href@noop {} {\bibfield  {journal} {\bibinfo  {journal} {unpublished}\ }
  (\bibinfo {year} {2012})}\BibitemShut {NoStop}%
\bibitem [{Note6()}]{Note6}%
  \BibitemOpen
  \bibinfo {note} {With tangential NBI heating, it is difficult to increase the
  toroidal Mach number --- and hence the equilibrium flow shear --- because of
  the fixed ratio of injected torque to power at a fixed injection
  energy.}\BibitemShut {Stop}%
\bibitem [{\citenamefont {Connor}\ \emph {et~al.}(1987)\citenamefont {Connor},
  \citenamefont {Cowley}, \citenamefont {Hastie},\ and\ \citenamefont
  {Pan}}]{connor_ppcf_1987}%
  \BibitemOpen
  \bibfield  {author} {\bibinfo {author} {\bibfnamefont {J.~W.}\ \bibnamefont
  {Connor}}, \bibinfo {author} {\bibfnamefont {S.~C.}\ \bibnamefont {Cowley}},
  \bibinfo {author} {\bibfnamefont {R.~J.}\ \bibnamefont {Hastie}}, \ and\
  \bibinfo {author} {\bibfnamefont {L.~R.}\ \bibnamefont {Pan}},\ }\href@noop
  {} {\bibfield  {journal} {\bibinfo  {journal} {Plasma Phys. Control. Fusion}\
  }\textbf {\bibinfo {volume} {29}},\ \bibinfo {pages} {919} (\bibinfo {year}
  {1987})}\BibitemShut {NoStop}%
\bibitem [{\citenamefont {Catto}\ \emph {et~al.}(1987)\citenamefont {Catto},
  \citenamefont {Bernstein},\ and\ \citenamefont {Tessarotto}}]{catto_pf_1987}%
  \BibitemOpen
  \bibfield  {author} {\bibinfo {author} {\bibfnamefont {P.~J.}\ \bibnamefont
  {Catto}}, \bibinfo {author} {\bibfnamefont {I.~B.}\ \bibnamefont
  {Bernstein}}, \ and\ \bibinfo {author} {\bibfnamefont {M.}~\bibnamefont
  {Tessarotto}},\ }\href@noop {} {\bibfield  {journal} {\bibinfo  {journal}
  {Phys. Fluids}\ }\textbf {\bibinfo {volume} {30}},\ \bibinfo {pages} {2784}
  (\bibinfo {year} {1987})}\BibitemShut {NoStop}%
\end{thebibliography}%
